\documentclass[12pt]{article}
\pdfoutput=1

\usepackage[english]{babel}
\usepackage{epsf,amssymb,amsmath,bbold,bbm}
\usepackage{graphicx,color}

\numberwithin{equation}{section} % restart equation  numbering in each section

\usepackage[colorlinks=true,       % false: boxed links; true: colored links
    linkcolor=blue,          % color of internal links (change box color with linkbordercolor)
    citecolor=blue,        % color of links to bibliography
    filecolor=blue,      % color of file links
    urlcolor=blue,   % color of external links
    linktoc=page	%defines which part of an entry in the table of contents is made into a link
     ]{hyperref}
\newcommand{\arXiv}[1]{\href{http://www.arXiv.org/abs/#1}{#1}}

\usepackage{cite}

\makeatletter
\renewcommand\section{\@startsection {section}{1}{\z@}%
                               {-3.5ex \@plus -1ex \@minus -.2ex}%nn
                               {2.3ex \@plus.2ex}%
                               {\normalfont\large\bfseries}}
\renewcommand\subsection{\@startsection{subsection}{2}{\z@}%
                                 {-3.25ex\@plus -1ex \@minus -.2ex}%
                                 {1.5ex \@plus .2ex}%
                                 {\normalfont\bfseries}}
\makeatother

\def\a{\alpha}

\def\d{\delta}

\def\eps{\epsilon}
\def\l{\lambda}

\def\G{\Gamma}

\def\del{\partial}

\def\Re{{\rm Re ~}}

\newcommand{\Id}{\mathbbm{1}}

\def\[{\left[}
\def\]{\right]}
\def\({\left (}
\def\){\right )}
\newcommand{\be}{\begin{equation}}
\newcommand{\ee}{\end{equation}}
\def\ba{\begin{eqnarray}}
\def\ea{\end{eqnarray}} 
\newcommand{\bea}{\begin{eqnarray}}
\newcommand{\eea}{\end{eqnarray}}
\newcommand{\beq} {\begin{equation}}
\newcommand{\eeq} {\end{equation}}

\def\ta{{\tilde \alpha}}

\newcommand{\bw}[0]{\bar{w}}
\newcommand{\bz}[0]{\bar{z}}
\newcommand{\bx}[0]{\bar{x}}

\newcommand{\bc}[0]{\bar{c}}

\newcommand{\hf}{{}_2F_{1} }
\newcommand{\hP}[1]{ h_{P_{#1}} }

\def\cx{{c_{x}}}

\begin{document}

\begin{titlepage}
%\begin{flushright}
%arXiv:
%\end{flushright}
%\vfill
\begin{center}
{\Large \bf  Heavy-Heavy-Light-Light correlators \\ \vskip 3mm  in Liouville theory}\\
 %
%\today

\vskip 10mm

{\large V.~Balasubramanian$^{a,b}$, A.~Bernamonti$^{c}$, B.~Craps$^b$,\\  T.~De Jonckheere$^b$, F.~Galli$^{c}$\\
\vspace{3mm}
}

\vskip 7mm

$^a$ David Rittenhouse Laboratory, Univ.~of Pennsylvania, 
 Philadelphia, PA 19104, USA \\
$^b$ Theoretische Natuurkunde, Vrije Universiteit Brussel, and \\ International Solvay Institutes,
Pleinlaan 2, B-1050 Brussels, Belgium \\
$^c$ Perimeter Institute for Theoretical Physics, \\
31 Caroline Street North, ON N2L 2Y5, Canada \\

\vskip 6mm
{\small\noindent  {\tt vijay@physics.upenn.edu, abernamonti@perimeterinstitute.ca, Ben.Craps@vub.ac.be, Tim.De.Jonckheere@vub.ac.be, fgalli@perimeterinstitute.ca}}

\end{center}
\vfill

\begin{center}
{\bf ABSTRACT}
\vspace{3mm}
\end{center}
We compute four-point functions of two heavy and two ``perturbatively heavy'' operators in the semiclassical limit of Liouville theory on the sphere.  We obtain these ``Heavy-Heavy-Light-Light'' (HHLL) correlators  to leading order in the conformal weights of the light insertions in two ways: (a) via a path integral approach, combining different methods to evaluate correlation functions from complex solutions for the Liouville field, and  (b) via the conformal block expansion. This latter approach identifies an integral over the continuum of normalizable states and a sum over an infinite tower of lighter discrete states, whose contribution we extract by analytically continuing standard results to our HHLL setting. The sum over this tower reproduces the sum over those complex saddlepoints of the path integral that contribute to the correlator. 
Our path integral computations reveal that when the two light operators are inserted at equal time in radial quantization, the leading-order HHLL correlator is independent of their separation, and more generally that at this order there is no short-distance singularity as the two light operators approach each other. The conformal block expansion likewise shows that in the discrete sum short-distance singularities are indeed absent for all intermediate states that contribute.  In particular, the Virasoro vacuum block, which would have been singular at short distances, is not exchanged.  The separation-independence of equal-time correlators is due to cancelations between the discrete contributions. 
These features lead to a Lorentzian singularity that, in conformal theories with anti-de Sitter (AdS) duals,  would be associated to locality below the AdS scale.

\end{titlepage}

\tableofcontents

%%%%%%%%%%%%%%%%%%%%%%%%%%%%%%%%%%%%%%%%%%%%%%%%%%%%%%%%%%%%%%%%%%%%%%%
%%%%%%%%%%%%%%%%%%%%%%%%%%%%%%%%%%%%%%%%%%%%%%%%%%%%%%%%%%%%%%%%%%%%%%%

\section{Introduction}
\label{sec:intro}

There has been significant recent interest in two-dimensional conformal field theories (CFTs) at large central charge. Via holography, these theories encode fundamental aspects of black hole physics, gravitational dynamics and the relation between geometry and entanglement. Much of the focus has been on the study of correlation functions via an expansion in conformal blocks. For 2d CFTs with a well-defined semiclassical limit and a sparse spectrum of low-dimension operators, semiclassical Virasoro conformal blocks have been used  to study aspects of black hole thermodynamics \cite{Fitzpatrick:2014vua,Fitzpatrick:2015zha}, holographic entanglement entropy \cite{Hartman:2013mia,Asplund:2014coa,Caputa:2014eta} and the information paradox \cite{Fitzpatrick:2016ive,Fitzpatrick:2016mjq}. Progress in understanding the structure of the block decomposition of 2d correlators at large $c$ was also achieved in the context of the D1-D5 system \cite{Galliani:2016cai}. Beyond the framework of holography, these techniques have been used to characterize 2d CFTs exhibiting scrambling of information \cite{Asplund:2015eha,deBoer:2016bov} and chaotic dynamics \cite{Roberts:2014ifa,Maldacena:2015waa,Perlmutter:2016pkf}, and to chart the space of 2d CFTs with conformal bootstrap techniques \cite{Chang:2015qfa,Collier:2017shs}.

An important technical ingredient in these works is that, under suitable assumptions about the CFT data,  it is possible at large central charge to approximate certain correlators by leading contributions in their conformal block expansion. Since contributions of individual Virasoro blocks are generically multi-valued, this raises the question of how exactly this multi-valuedness recombines into a single-valued correlator. We address this question explicitly in Liouville theory, a CFT which has a well-defined semiclassical limit and whose spectrum and OPE coefficients are explicitly known, allowing us to compute correlators via path integrals as well as via conformal block expansions.

Specifically, in the semiclassical limit, we compute four-point functions in Liouville theory on the sphere of two heavy and two ``perturbatively heavy'' operators. All four operators have conformal dimensions scaling as the central charge, with the prefactor being perturbatively small for the ``perturbatively heavy'' operators. Following terminology from holographic CFTs, we will refer to the latter operators as ``light'', even though in Liouville theory the terminology ``light'' has traditionally been reserved for operators with  fixed conformal dimension in the semiclassical limit. So we refer to the correlation functions of interest as Heavy-Heavy-Light-Light (HHLL) correlators. 

In the path integral approach to semiclassical Liouville field theory, correlation functions involving heavy (and perturbatively heavy) operators are obtained in a saddlepoint approximation starting from solutions of the Liouville equation with $\d$-function sources at the locations of the heavy insertions. 
Correlators of heavy and perturbatively heavy operators were considered on the sphere \cite{Menotti:2004xz}, on the pseudosphere \cite{Menotti:2005fk,Menotti:2006gc} and in the conformal boundary case \cite{Menotti:2006tc} in instances in which the path integral is dominated by a unique real saddlepoint. However, as we will review, no real saddlepoint exists for HHLL correlators in Liouville theory on the sphere, and these four-point functions have not been computed before. The progress reported here has been possible thanks to the modern interpretation of complex saddles of \cite{Harlow:2011ny}.   We will work to leading order in the semiclassical limit and in a linearized approximation in the conformal weight of the light operators, and do not attempt to compute perturbative corrections in powers of the Liouville coupling constant $b$.

We use a constructive approach that combines  three different methods.  First, after setting the scene in Sec.~\ref{sec:review},  in Sec.~\ref{sec:linearized}  we superpose complex Liouville field saddlepoints for three-point functions  \cite{Harlow:2011ny} to compute the four-point function.
 This approach has a limited regime of validity, as it only holds for specific values of the cross ratio of the four insertion points. There is a countable infinity of complex saddles; the corresponding on-shell actions have constant imaginary part and differ from one another by multiples of $2\pi i$.

Second, in Sec.~\ref{sec:mono} we use the monodromy method to solve for the functional dependence of the saddles in the whole complex plane. Previous work considered the regime in which a unique real and single-valued saddle was guaranteed to exist, see e.g. \cite{Zamolodchikov:1995aa,Menotti:2004xz,Hadasz:2006rb}. This corresponds to a situation in which the sum of conformal weights is above a certain threshold, in which case one has to impose SU(1,1) monodromy. For our HHLL correlator there is no real saddle. However, the results of Sec.~\ref{sec:linearized} lead us to recast the computation of the complex saddles into an SU(2) monodromy problem for the real part of the Liouville field. We are then able to compute the full functional dependence of the action for saddlepoint contributions and find a universal dependence on the insertion points. This implies that the sum over all contributing saddles, which we will  determine by other methods, will share this same dependence on the insertion points.

Third, in Sec.~\ref{sec:DOZZ} we calculate the four-point function via a decomposition into conformal blocks.  This procedure is subtle in Liouville theory, which has a continuum of normalizable states. Analytic continuation of classic results to our HHLL setting reveals the additional contribution of an infinite tower of discrete, non-normalizable intermediate states. The corresponding tower of conformal weights starts at twice the weight of the light operators in the correlator.  Higher weights in the tower are shifted by  non-negative integers $m$. Summing over these discrete intermediate states leads to a single-valued result, which reproduces by itself the outcome of the path integral calculation of Sec.~\ref{sec:linearized} and \ref{sec:mono}. Moreover, this computation also determines the set of saddlepoints that contribute to the correlator.  In this way, comparison between the path integral approach and the block decomposition reveals that, in HHLL Liouville correlators, single-valuedness is recovered from recombination of discrete exchanges alone.

In the standard approach for computing correlation functions of heavy operators and operators that are light in the Liouville sense (as opposed to perturbatively heavy, as discussed above) one treats light operators as probes in the background produced by the heavy insertions, and integrates over the moduli of this field configuration. As we discuss in Sec.~\ref{sec:discussion}, one could expect this approach extends to perturbatively heavy insertions at the linearized level.
This is not a priori obvious, as for HHLL insertions there is a complex modulus one needs to integrate over. We do not perform here a full analysis, but notice that assuming specific saddles dominate the integration over the modulus, one in fact reproduces the result obtained via the previous methods.

The relation between Liouville field theory and pure 3d gravity in anti-de Sitter space has been long debated, see for instance \cite{Krasnov:2000ia} and references therein. Important points in this discussion include the non-normalizability of the Liouville vacuum and the fact that the spectrum of normalizable states has a gap, and thus does not by itself lead to the expected asymptotic growth of states.  Despite these puzzles, as we discuss in Sec.~\ref{sec:discussion}, we find that the structure of the HHLL correlator  that we study displays  similarities to holographic CFTs. This can be traced back to the fact that the block decomposition contains discrete contributions, which resum into a single-valued result with no need to take into account the contribution of the continuum part of the spectrum. One similarity is that the resummation we find is that of ``double-trace'' exchanges, which also occurs in holographic CFTs. A particularly intriguing feature is that our HHLL Liouville correlator also exhibits a Lorentzian singularity that,  in conformal theories with AdS duals, would be associated to locality below the AdS scale \cite{Gary:2009ae,Heemskerk:2009pn}. 

We finish the paper with remarks on the fact that  certain HHLL correlators in holographic CFTs compute the single interval entanglement entropy in excited states \cite{Asplund:2014coa}. We comment on the possibility of interpreting our result from this perspective.

%%%%%%%%%%%%%%%%%%%%%%%%%%%%%%%%%%%%%%%%%%%%%%%%%%%%%%%%%%%%%%%%%%%%%%%
%%%%%%%%%%%%%%%%%%%%%%%%%%%%%%%%%%%%%%%%%%%%%%%%%%%%%%%%%%%%%%%%%%%%%%%

\section{Semiclassical correlators in Liouville theory}
\label{sec:review}

In this section we introduce and review the minimal set of information needed for our analysis and computations. We focus on those general aspects that are important for computing correlators of heavy and perturbatively heavy operators in the semiclassical limit via a path integral approach.  This will be relevant for Secs.~\ref{sec:linearized} and~\ref{sec:mono}.  We postpone the discussion of the conformal block decomposition and of the well-known DOZZ three-point function coefficients to Sec.~\ref{sec:DOZZ}. Comprehensive reviews of Liouville theory and of its applications include for instance \cite{Seiberg:1990eb,ZZ,Teschner:2001rv}.

%%%%%%%%%%%%%%%%%%%%%%%%%%%%%%%%%%%%%%%%%%%%%%%%%%%%%%%%%%%%%%%%%%%%%%%

\subsection{Action, Liouville equation and correlators in the semiclassical limit}

Liouville theory defines a conformal algebra with central charge 
\be
c = 1 + 6 Q^2 = 1+ 6 \(b + \frac 1 b\)^2\,. 
\label{eq:Qdef}
\ee
The CFT data of this 2d conformal field theory are known explicitly. 
Primary operators are exponentials 
\be
 V_{\alpha} =  e^{2 \alpha \phi}\, ,
\ee
where $\phi$ is the Liouville field and  $\a$  is known as the Liouville momentum.  These operators have conformal weights
\be
h_{\alpha }= \bar h_{\alpha} = \alpha(Q-\alpha)\, .  \label{eq:halpha0}
\ee
Vertex operators with momenta $\a$ and $Q-\a$ have equal weights and they are interpreted to correspond to the same quantum operator up to a rescaling \cite{Seiberg:1990eb}. The three-point function coefficients are also known and given by the famous DOZZ formula \cite{Dorn:1994xn,Zamolodchikov:1995aa}, which we will give explicitly in Sec.~\ref{sec:DOZZ}.

The  field theory is governed by the Liouville action 
\be
S  = \frac{1}{4\pi} \int d^2x~ \sqrt{\hat g}~\[ \hat g^{ab} \del_a \phi \, \del_b \phi + Q \hat R \phi + 4 \pi  \mu e^{2 b \phi}\]
\ee
with $\mu>0$.  
On the two-sphere, it is customary to take $\hat g_{ab}$ to be the flat metric $d\hat s^2= dz d\bz$ with the asymptotic condition%
\footnote{This allows for a smooth ``physical metric'' $g_{ab} = e^{\frac 2 Q \phi} \hat g_{ab}$ on the two-sphere (see e.g.\ \cite{Harlow:2011ny} for a more detailed discussion).}
\be \label{eq:asympt}
\phi(z,\bz) = - 2 Q \log |z|  + O(1)\,, \qquad |z|\to \infty.
\ee
The semiclassical limit corresponds to sending $b \to0$. This is more conveniently studied  using the rescaled classical field $\phi_c \equiv 2 b \phi$. The action expressed in terms of $\phi_c$ scales  like $b^{-2}$ and the Liouville equation descending from its variation reads
\be
\del_z \del_{\bz} \phi_c = 2 \l \, e^{\phi_c} \,,  \label{eq:Lsphere0}
\ee
where 
\be
\lambda \equiv \pi \mu b^2
\label{eq:lambdadef}
\ee
is held fixed in the limit $b\to 0$.  
The asymptotic condition \eqref{eq:asympt} becomes
\be  \label{eq:BC}
\phi_c(z,\bz) = - 4 \log |z| + O(1)\,, \qquad |z|\to \infty \,.
\ee

When considering correlation functions of primary fields
 \be
\langle V_{\a_1}(z_1,\bz_1) V_{\a_2}(z_2,\bz_2) V_{\a_3}(z_3,\bz_3)\dots   \rangle  = \int \mathcal{D}[\phi_c] e^{-S[\phi_c]}\prod_{i} e^{\frac{\a_i}{b}\,  \phi_c(z_i,\bz_i)} \label{eq:Lpathint}
\ee
in the semiclassical limit there are two main classes of operators that can be identified depending on the scaling of $\a_i$ with $b$: ``heavy" operators, with momentum $\alpha = \eta/b$ and $\eta$ held fixed as $b\to0$, and  ``light" operators with $\a = b \, \sigma$ and $\sigma$ fixed as $b\to0$.  The first scale as the action itself and therefore have a non-trivial effect  in determining the classical solutions, while the latter do not affect the saddlepoints. In terms of conformal weights, a heavy insertion has 
\be
h_{\alpha }= \bar h_{\alpha} = \frac{\eta(1-\eta)}{b^2} +O(1) \, .  \label{eq:halpha}
\ee
The effect of heavy insertions is to modify the classical Liouville equation \eqref{eq:Lsphere0}  by $\delta$-function terms: 
 \be  \label{eq:Lsphere}
\del_z \del_{\bz} \phi_c  =  2 \lambda \, e^{\phi_c}-   2 \pi \sum_i \eta_i  \delta^{(2)}(z-z_i) \,.
\ee
In view  of the equivalence between momenta explained just below \eqref{eq:halpha0}, we can assume that these heavy operators satisfy Re$(\eta_i)< 1/2$, which is known as the Seiberg bound \cite{Seiberg:1990eb}. 
In a neighbourhood of each insertion we then have the behaviour
\be \label{eq:singular}
\phi_c(z,\bz) = - 4\eta_i \log |z - z_i| + O(1)\,, \qquad |z - z_i |\to 0 \,.
\ee
In fact, divergences in the evaluation of the path integral are introduced both by taking the flat reference metric $\hat g_{ab}$ (see \eqref{eq:BC}), and by inserting heavy operators (see \eqref{eq:singular}).  They can be systematically regularized, giving rise to a  modified action $\tilde S$ that takes into account the presence of the  heavy insertions  
\cite{Zamolodchikov:1995aa} (see also \cite{ZZ,Harlow:2011ny}). Without entering into the details of this procedure, which will not be needed here, the expectation value of a correlator of the form \eqref{eq:Lpathint}, involving heavy and light insertions (in the standard Liouville sense), can then be approximated as
\be
\langle  V_\frac{\eta_1}{b}(z_1,\bar{z}_1) \dots V_\frac{\eta_j}{b}(z_j,\bar{z}_j) V_{b \sigma_1}(w_1,\bar{w}_1)\dots V_{b \sigma_n}(w_n,\bar{w}_n) \rangle \approx
e^{-\frac{\tilde{S}[\phi_c] }{b^2}} \prod_{k=1}^{n} e^{ \sigma_{k} \phi_c(w_k,\bar{w}_k) } \,  .  \label{eq:Lsemi}
\ee
In writing the expression on the right hand side, we have explicitly isolated and extracted  the scaling factor $b^{-2} \approx c/ 6$ from the renormalized  semiclassical action $ \tilde{S}[\phi_c] $. The latter is evaluated on a solution to the equation of motion \eqref{eq:Lsphere}. In general  there will be several contributing saddles, and in such a case the right hand side \eqref{eq:Lsemi} is understood to involve a sum over them. 

We will actually not need the explicit functional form of the action $ \tilde{S}  $ in order to evaluate it on a solution of the Liouville equation \eqref{eq:Lsphere} with boundary condition \eqref{eq:BC}. 
In fact, given a classical solution, the on-shell action  can be obtained from known relations between the behavior of the Liouville field $\phi_c$ near the heavy insertion points and derivatives of $\tilde S$. The semiclassical solution has the asymptotics
\be \label{eq:phiasymptotic}
\phi_c(z, \bar z) \approx -4\eta_i \log |z-z_i| + \sigma_i - \frac{c_i}{\eta_i} (z-z_i) - \frac{\bar c_i}{\eta_i} (\bar z- \bar z_i) + \dots \,, \qquad \quad |z -  z_i| \to 0 
\ee
near any heavy insertion at $(z_i, \bz_i)$, with the leading logarithm corresponding to the singularity required by the heavy operator, as in \eqref{eq:singular}. 

The first differential relation we will use, relates the constant term in \eqref{eq:phiasymptotic} to the derivative of the action with respect to the momenta of the heavy insertions \cite{Zamolodchikov:1995aa}
\be \label{eq:intsigma}
\frac{\del \tilde  S  }{\del \eta_i} = -\sigma_i\,.
\ee
This can be integrated to obtain the correlator with heavy insertions up to a $\eta_i$-independent integration constant. Notice that if there are multiple insertions  with the same $\eta_i$, on the right hand side one has to sum over the  corresponding $\sigma_i$.

A similar  differential condition, known as the Polyakov relation,  involves the coefficient of the linear term in the expansion
\be \label{eq:intC}
\frac{\del \tilde  S }{\del z_i} = c_i \, , 
\ee
and similarly for $\bz_i$ and $\bar c_i$, which is the conjugate of $c_i$.  The parameters $c_i$ are  known as accessory parameters and are non-holomorphic  functions, as we will review in Sec.~\ref{sec:mono}. 

%%%%%%%%%%%%%%%%%%%%%%%%%%%%%%%%%%%%%%%%%%%%%%%%%%%%%%%%%%%%%

\subsection{Complex solutions for two and three heavy insertions}

As we will discuss  in Sec.~\ref{sec:mono}, for real $\eta_i<1/2$ and 
$\lambda>0$, a unique real, single-valued solution to the Liouville equation  \eqref{eq:Lsphere} 
exists if and only if $\sum_i \eta_i >1$.  Conversely, if $\sum_i \eta_i <1$, a unique real and single valued solution would exist only for 
$\lambda<0$ (see for instance \cite{ZZ}). 

More generally, we will be interested in solutions that do not satisfy the above constraints on the sum of the $\eta_i$. We will therefore drop the reality condition and allow for complex solutions. 
This approach was first explored in \cite{Harlow:2011ny}, where general complex solutions  of the Liouville equation for two and three heavy insertions (HH and HHH respectively) were found using a standard procedure, which we will also discuss and apply in Sec.~\ref{sec:mono}. 

Before reviewing the results of \cite{Harlow:2011ny} for HH and HHH correlators, there are two specific aspects that are important for our analysis. First, notice that the classical Liouville equation is invariant under the shift $\phi_c \to \phi_c +  2 \pi i N$, with $N$ an integer. The multiple complex solutions found in \cite{Harlow:2011ny}  are indeed related to one another by this kind of shift.  Second, known results on two and three-point functions for real $\eta_i$ are recovered and can be reinterpreted in terms of sums over semiclassical complex solutions of this type  \cite{Harlow:2011ny}. \\

For two heavy insertions with equal conformal weights, 
the general complex semiclassical solution for the Liouville field reads  \cite{Harlow:2011ny} 
\be \label{eq:phiHW}
e^{\phi_c (z, \bar z)} = \frac{1}{\lambda} \left[ \kappa |z- z_1|^{2 \eta} |z -z_2|^{2-2\eta} - \frac{1}{\kappa (1-2 \eta)^2|z_{12}|^2} |z-z_1|^{2 - 2 \eta} |z- z_2|^{2\eta} \right]^{-2}\, .
\ee
Here $\kappa$ is a complex number, constrained by requiring the absence of additional singularities other than \eqref{eq:BC} and  \eqref{eq:singular}. For $\eta$ real, $\kappa$ takes values in the upper half-plane with the real axis removed. 
Moreover, as mentioned above, although this gives the general form of $e^{\phi_c}$, when evaluating  $\phi_c$ itself there is the extra freedom of choosing the branch of the logarithm.

The corresponding on-shell action is independent of $\kappa$, as one can check explicitly from the constant terms $\sigma_i$ in the field expansions around the insertions points:
\begin{align}
\sigma_1& = 2 \pi i \(N + \frac{1}{2}\) - \log \lambda - 2 \log( -i \kappa) + (4 \eta- 4)\log |z_{12}| \, ,    \\
\sigma_2& =  2 \pi i \(N + \frac{1}{2}\) - \log \lambda + 2 \log( -i \kappa) + 4 \eta \log |z_{12}|  +   4 \log (1- 2 \eta) \, ,
\end{align}
where the integer $N$ corresponds to the choice of branch in the logarithm. 
Then,  according to  \eqref{eq:intsigma}, 
\be\label{diffeq}
\frac{\del \tilde  S  }{\del \eta} = -\sigma_1 - \sigma_2
\ee
and it is immediate to check that the dependence on $\kappa$ simply cancels out. The differential equation \eqref{diffeq} determines the on-shell action, and thus the two-point function up to constant terms in $\eta$. In \cite{Harlow:2011ny} these were fixed by requiring  consistency with the limiting case  $\eta=0$.
The form of the action obtained in this way is
\begin{align}
\tilde{S}_N  =& - (1-2\eta) \log \lambda + 2\pi i \(N+\frac{1}{2}\) ( 1-2\eta ) + 2 (1-2\eta)  \[ \log(1-2\eta) - 1 \] \nonumber \\
& +4\eta(1-\eta)\log |z_{12} |  \, .    \label{eq:2ptaction}
\end{align}
We will use this result in Sec.~\ref{sec:linearized} to fix an undetermined constant when evaluating the HHLL correlator.
The action \eqref{eq:2ptaction} corresponds to a single saddle out of a family of solutions parametrized by $\kappa$ and integers $N$.  Summing over a particular infinite set of such saddles and interpreting the divergent integral over the modulus $\kappa$ appropriately, one reproduces the two-point function $ \langle V_{\eta/b}(z_1,\bz_1) V_{\eta/b}(z_2,\bz_2) \rangle$ obtained from analytic continuation of the DOZZ formula  \cite{Harlow:2011ny}.
 
The general complex semiclassical solution for the Liouville field in the presence of three heavy insertions is the analytic continuation in $\eta_i$ of the real solution of  \cite{Zamolodchikov:1995aa}, and was worked out in \cite{Harlow:2011ny}. It reads
\be\label{eq:ThreePoint}
e^{\phi_c (z,\bz)} = \frac{1}{\lambda} \frac{ \left| z-z_2\right|^{-4}}{ \[ a_1 P^{\eta_1}(w) P^{\eta_1}(\bar{w}) - a_2 P^{1-\eta_1}(w) 
P^{1-\eta_1}(\bar{w})\]^2} \, ,
\ee
where the functions  $P^{\eta}$ are   Riemann functions
\begin{align}
&P^{\eta_1} (w) = w^{\eta_1} (1-w)^{\eta_3} ~{}_2F_1\[\eta_1 + \eta_3-\eta_2, \eta_1+\eta_2+\eta_3 -1, 2\eta_1, w\] \, , \\
&P^{1-\eta_1}(w)= w^{1-\eta_1} (1-w)^{1-\eta_3} ~{}_2F_1\[1-\eta_1+\eta_2-\eta_3,2-\eta_1-\eta_2-\eta_3,2-2\eta_1,w\] \, .
\end{align}
The parameters $a_1$ and $a_2$ are determined, up to an irrelevant choice of sign,  through the conditions
 \begin{align}
a_1a_2 &= \frac{\left|z_{13}\right|^2}{\left|z_{12}\right|^2 \left| z_{23}\right|^2 (1-2\eta_1)^2},\label{eq:a1a2}\\
a_1^2 &=  \frac{\left|z_{13}\right|^2}{\left|z_{12}\right|^2 \left| z_{23}\right|^2} \frac{ \gamma(\eta_1 + \eta_2 -\eta_3)\gamma(\eta_1 + \eta_3 -\eta_2)\gamma(\eta_1 + \eta_2 +\eta_3-1)}{\gamma(\eta_2 + \eta_3 -\eta_1)\gamma(2\eta_1)^2},\label{eq:a1}
\end{align}
and we have defined the combinations 
\be
\gamma (w) \equiv \frac{\Gamma(w) }{ \Gamma(1-w)} \, ,\qquad \qquad  w\equiv \frac{ (z-z_1)z_{23}}{(z-z_2)z_{13} }  \, . 
\ee
As discussed in \cite{Harlow:2011ny}, depending  on the values of $\eta_i$, there are various subtleties  in requiring that the solution has no additional singularities other than those corresponding to  the heavy insertions.  The range of parameters we will consider is that of real $\eta_i$. From \eqref{eq:a1a2} and \eqref{eq:a1} it follows that $a_1$ and $a_2$ are purely imaginary and therefore no cancellations occur in the expression in the denominator of   \eqref{eq:ThreePoint}.\footnote{This is also the case when momenta $\eta_i$ have a small imaginary part.  The region of parameters  we will be considering in  the next sections corresponds to what is indicated as Region II in \cite{Harlow:2011ny}.} 
As in the case of  the two-point function,  \cite{Harlow:2011ny}  showed how  the general complex  saddles reproduce results obtained directly from the DOZZ formula, extending the analysis of \cite {Zamolodchikov:1995aa}.

The Liouville solution in the case of four heavy insertions and the corresponding semiclassical correlator are not known explicitly. Here we are not able to compute such a correlator in full generality.  But as in the case of semiclassical conformal Virasoro blocks  \cite{Hartman:2013mia,Fitzpatrick:2014vua,Fitzpatrick:2015dlt,Beccaria:2015shq,Fitzpatrick:2015zha},  
progress can be made  in the limit where some of the heavy operators have a perturbatively small coefficient $\eta$.

We consider a four-point function of heavy operators with $\a_1=\a_2=\a_H \equiv \eta_H/b$ and $\a_3=\a_4=\a_L\equiv \eta_L/b$ with $\eta_L \ll 1$. The notation $\eta_H$ and $\eta_L$ is to distinguish between insertions with $\eta_H \sim O(1)$, which we refer to as \textit{heavy}, and perturbatively heavy insertions with $\eta_L \ll 1$, which with a slight abuse of terminology we refer to as \textit{light}. Light operators in the standard Liouville sense will not appear in the following sections. 
This choice of momenta corresponds to the Heavy-Heavy-Light-Light correlator
\be \label{eq:HHLL}
 \langle V_{\frac{\eta_H}{b}}(z_1,\bz_1) V_{\frac{\eta_H}{b}}(z_2,\bz_2) V_{\frac{\eta_L}{b}}(z_3,\bz_3) V_{\frac{\eta_L}{b}}(z_4,\bz_4)  \rangle \,.
\ee
This is a four-point function of primaries with conformal weights that scale as 
\begin{align}
\frac{h_H}{c}  \sim \eta_H (1-\eta_H),  \qquad \quad  \frac{h_L}{c}   \sim\eta_L\ll1 
\end{align}
in terms of the central charge $c \approx 6/b^2 \to \infty$.

%%%%%%%%%%%%%%%%%%%%%%%%%%%%%%%%%%%%%%%%%%%%%%%%%%%%%%%%%%%%%%%%%%%%%%%%
%%%%%%%%%%%%%%%%%%%%%%%%%%%%%%%%%%%%%%%%%%%%%%%%%%%%%%%%%%%%%%%%%%%%%%%%
\section{Superposition of linearized three-point functions}
\label{sec:linearized}

In this section we compute the on-shell action perturbatively in an expansion in  $\eta_L \ll 1$ using the relation \eqref{eq:intsigma}.  To do this, we need to work out the expansion of the semiclassical Liouville field for HHLL insertions and read out the constant terms $\sigma_i$ in the $z\to z_i$ limit. The basic idea is that, when working at linear order in $\eta_L$ and knowing the full solution \eqref{eq:ThreePoint} for two heavy and one light (perturbatively heavy) insertions, one can linearly superpose the effect of the two light insertions. At zeroth order in $\eta_L$, the light operators can be neglected completely and the solution for the Liouville field coincides with the two-point solution \eqref{eq:phiHW} for two heavy operators at $z_1$ and $z_2$. At first order, the two light operators act as perturbatively small sources that modify the stress tensor and therefore the solution of the Liouville field. At this order, one can superpose the effects of the insertions at $z_3$ and $z_4$, just as in electrodynamics. That is, we can consider the three-point solution $\phi_c^{(123)}$ with insertions at $z_1, z_2$ and $z_3$, linearized in the Liouville momentum of a perturbatively light insertion at $z_3$
\be \label{eq:linearizedHHH}
\phi_c^{(123)} \approx \phi_c^{(12)} +\eta_L \varphi_c^{(3)} + O(\eta_L^2) \, . 
\ee
The zeroth order contribution is simply the solution for two heavy insertions $\phi_c^{(12)}$, while the linear contribution $\varphi_c^{(3)}$ accounts for the presence of the third operator. Similarly, the solution $\phi_c^{(124)}$ with insertions at $z_1,   z_2$ and $z_4$ will provide a linear contribution describing the effect of the operator at $z_4$. Up to linear order, the Liouville field is then given by
\begin{equation} \label{eq:linearizedsol}
\phi_c \approx \phi_c^{(12)} + \eta_L \varphi_c^{(3)} + \eta_L \varphi_c^{(4)} + O(\eta_L^2) \, .  
\end{equation}

Above, we glossed over an important subtlety.  When we obtain a two-point solution $\phi_c^{(12)}$ as   the zeroth order in the $\eta_L$ expansion of the three-point solution, the result depends on the location of the third insertion.   This will restrict the applicability of our method to cases where the two light insertions are on the same ``circle of Apollonius," as we will now explain.
Consider the solution for three heavy insertions given by \eqref{eq:ThreePoint}-\eqref{eq:a1} in Sec.~\ref{sec:review}. When the insertions at $z_1$ and $z_2$ have the same weight with $\eta_1=\eta_2=\eta_H$ and  we set to zero the weight $\eta_3=\eta_L$ of the third insertion at  $z_3$, we  have
\begin{equation} \label{eq:TwoPointSolution}
e^{\phi_c^{(123 )}} =  \frac{1}{\lambda} \[ \frac{i | z- z_1|^{2\eta_H}  | z- z_2 |^{2-2\eta_H} | z_{23}|^{2\eta_H-1}}{(1-2\eta_H ) |z_{12}|  |z_{13}|^{2\eta_H-1} }- \frac{ | z- z_1|^{2-2\eta_H} | z-z_2|^{2\eta_H}\left| z_{13}\right|^{2\eta_H - 1}}{i (1-2\eta_H )  |z_{12}|  |z_{23} |^{2\eta_H-1} }\]^{-2} \,.
\end{equation}
Comparing with the general complex two-point solution  \eqref{eq:phiHW}
\begin{equation}
e^{\phi_c^{(12)}} = \frac{1}{\lambda} \[ \kappa \left| z- z_1\right|^{2\eta_H} \left| z- z_2\right|^{2-2\eta_H} - \frac{1}{\kappa (1-2\eta_H)^2 | z_{12} |^2} \left| z- z_1\right|^{2-2\eta_H}\left| z-z_2\right|^{2\eta_H} \]^{-2} 
\label{eq:TwoPointSolution2}
\end{equation}
we see that we must take 
\begin{equation} \label{eq:2ptconst}
\kappa = \frac{i}{(1-2\eta_H )  \left|z_{12}\right|} \left(\frac{\left| z_{13}\right|}{ \left|z_{23}\right|}\right)^{1-2\eta_H} \,.
\end{equation}
That is, analytically continuing in the weight of the  light insertion at $z_3$ selects a specific $\kappa$ in the two-point solution, which depends non-trivially on the light insertion point.

Similarly, we could have started with the three-point solution $\phi_c^{(124)}$, which amounts to replacing $z_3\rightarrow z_4$. 
Clearly we can only do perturbation theory in $\eta_L$ if the zeroth order solution of $\phi_c^{(123)}$ agrees with the zeroth order solution of $\phi_c^{(124)}$.  Therefore, we get a non-trivial condition for the validity of the linearized construction,
\begin{equation}\label{eq:Apollonius}
\frac{\left|z_{14}\right|}{\left| z_{24}\right|} =\frac{\left|z_{13}\right|}{\left| z_{23}\right|}\,.
\end{equation}
This is precisely the condition that $z_3$ and $z_4$ lie on a ``circle of Apollonius''.\footnote{This is an ancient criterion for defining points on a circle named after Apollonius of Perga (262 BC-190 BC), a Greek geometer and astronomer. He is most famous for his work on conic sections and  named the hyperbola, ellipse and parabola  \cite{Appolonius}.} While this is an important restriction, it does capture an interesting special case: if one inserts the heavy operators at zero and infinity in order to create initial and final states in radial quantization, then circles of Apollonius are equal-time circles around the origin. So our method will allow us to compute correlators with light operators inserted at equal times.  It will also be instrumental for carrying out the computation in Sec.~\ref{sec:mono}, where this restriction will be removed.  

%%%%%%%%%%%%%%%%%%%%%%%%%%%%%%%%%%%%%%%%%%%%%%%%%%
\subsection{Computation of the on-shell action with equal-time light insertions}\label{subsec:computation}

When expanding the Liouville HHLL solution around the light insertions to use the relation \eqref{eq:intsigma}, here 
\begin{equation} \label{eq:sigma34}
\frac{d\tilde{S}}{d\eta_L} = - \sigma_3  -\sigma_4
\end{equation}
the integration is in $\eta_L$. Therefore the first order action is obtained directly from the constant term at  zeroth order of the Liouville field, which  comes entirely from the two-point solution above. 
The relevant asymptotic expansions at zeroth order in $\eta_L$ near the light insertions are
\begin{align}
\phi_{c,N}(z\rightarrow z_3) &=   -\log\lambda +2\pi i \(N + \frac{1}{2}\) + 2\log ( 1- 2\eta_H ) - 2\log 2 - 2 \log \frac{ \left| z_{13}\right| \left|z_{23}\right|}{\left| z_{12}\right|}  +   \nonumber\\
& +\( \frac{1}{z_{13}}+\frac{1}{z_{23}}\) (z -z_3) + \dots  \label{eq:asym3} \\
\phi_{c,N}(z\rightarrow z_4) &=   -\log\lambda +2\pi i \(N + \frac{1}{2}\) + 2\log (1- 2\eta_H ) - 2\log 2 - 2 \log \frac{ \left| z_{14}\right| \left|z_{24}\right|}{\left| z_{12}\right|} +   \nonumber\\
& +\( \frac{1}{z_{14}}+\frac{1}{z_{24}}\) (z -z_4) + \dots \, . \label{eq:asym4}
\end{align}
 The integer $N$  arises from taking the logarithm of \eqref{eq:TwoPointSolution}, as in Sec.~\ref{sec:review}. It effectively labels a countably infinite number of complex saddles $\phi_{c,N}$  for the four-point function, which are related one to the other by $2\pi i $ shifts. 

From \eqref{eq:sigma34} we find
\begin{align} \label{eq:action}
\tilde{S}{_N}  &= \tilde{S}_N^{(0)} + 2\eta_L \log \lambda - 4\pi i \(N+\frac 1 2\)\eta_L  - 4\eta_L \log (1-2\eta_H) + 4\eta_L\log 2\nonumber\\
& + 2\eta_L   \log \( \frac{ | z_{13}| |z_{14}|| z_{23}| | z_{24}| }{ |z_{12}|^2} \) + \dots 
\end{align} 
where we denoted  the $\eta_L$-independent part as $\tilde{S}_N^{(0)}$. The latter is just determined from the two-point solution for heavy insertions and was given in  \eqref{eq:2ptaction}
\begin{align} \label{eq:St0}
\tilde{S}_N^{(0)} =& - (1-2\eta_H) \log \lambda + 2\pi i \(N+\frac{1}{2}\) ( 1-2\eta_H) + 2 (1-2\eta_H)  \[ \log(1-2\eta_H) - 1 \] \nonumber \\
& +4\eta_H(1-\eta_H)\log |z_{12} |  \,. 
\end{align}
Altogether we arrive at the final expression
\begin{align} 
\tilde{S}{_N}  &= - (1- 2\eta_H- 2\eta_L) \log \lambda + 2\pi i \(N+\frac{1}{2}\) ( 1-2\eta_H - 2\eta_L) - 2 (1-2\eta_H) \nonumber \\
& + 2 (1-2\eta_H - 2\eta_L) \log(1-2\eta_H)   + 4\eta_L\log 2   +4\eta_H(1-\eta_H)\log |z_{12} |  \nonumber \\ 
& + 2\eta_L   \log \( \frac{ | z_{13}| |z_{14}|| z_{23}| | z_{24}| }{ |z_{12}|^2} \) + \dots \label{eq:Sconst}
\end{align} 
where we recall that this result only holds for insertions satisfying the ``circle of Apollonius'' relation \eqref{eq:Apollonius}.  
This is the main result of this section, which shows that the saddles are generically complex and related to one  another by shifts proportional to $ 2 \pi i $. It also shows that in radial quantization for light insertions at equal time, there is no dependence on the angular separation. 
Below we collect some results useful for checking the consistency of our solution.

%%%%%%%%%%%%%%%%%%%%%%%%%%%%%%%%%%%%%%%
\subsection{Consistency checks}\label{subsec:consistency}

An alternative to the approach above would have been to use \eqref{eq:intsigma} with an expansion around the heavy insertions
\begin{equation} \label{eq:sigma12}
\frac{d\tilde{S}}{d\eta_H} = - \sigma_1 -\sigma_2\, .
\end{equation}
This requires knowledge of  the Liouville field up to first order in $\eta_L$. 
The HHL solution reads
\be
e^{\phi_c^{(123)}} = \frac{1}{\lambda} \frac{ \left| z-z_2\right|^{-4}}{ \[ a_1 P^{\eta_H}(w) P^{\eta_H}(\bar{w}) - a_2 P^{1-\eta_H}(w) 
P^{1-\eta_H}(\bar{w})\]^2}
\ee
with
\begin{align}
w&= \frac{z_{23} (z-z_1)}{z_{13}(z-z_2)} 
\end{align}
and to linear order in  $\eta_L$\footnote{More precisely, to have a well defined perturbative expansion in $\eta_L$, in working at fixed $\eta_H<1/2$ and taking the small $\eta_L$ limit we are additionally always assuming  that $\eta_L \ll 1- 2 \eta_H$.}
\begin{align}
a_1  &\approx\pm  i \frac{\left|z_{13}\right|}{\left|z_{12}\right| \left| z_{23}\right| (1-2\eta_H)} \(1 + \frac{\eta_L}{1-2\eta_H}  \)\\
a_2  &\approx \mp i \frac{\left|z_{13}\right|}{\left|z_{12}\right| \left| z_{23}\right| (1-2\eta_H)} \(1 - \frac{\eta_L}{1-2\eta_H} \)\,.
\end{align}
In the proximity of the heavy insertion at $z_1$ ($w \to 0$), the  Riemann $P^{\eta}$-function is
\begin{align}
P^{\eta_H} (w) &\approx w^{\eta_H} \(1- \frac{w}{2 \eta_H}  \eta_L\)  \, .
%P^{1-\eta_H}(w) &\approx w^{1-\eta_H}  \(1- \frac{w}{2 (1-\eta_H)} \eta_L\) \,.
\end{align}
These give
\begin{align}
& \phi_{c,N}^{(123)}(z\to z_1) = - 4 \eta_H \log |z-z_1| -\log\lambda +2\pi i \(N + \frac{1}{2}\) + 2\log(1- 2\eta_H)  \label{eq:asym123} \\
 &+ 2(1- 2\eta_H) \log \frac{ | z_{23}  |}{ |z_{13} | | z_{12} |}   -\frac{2}{(1-2 \eta_H)}  \eta_L  -  \frac{2(1-\eta_H)}{z_{12}} (z -z_1)  +  \frac{\eta_L}{ \eta_{H}} \frac{z_{23}}{z_{12} z_{13} }  (z -z_1) +\dots       \nonumber
\end{align}
and an analogous expression for  $\phi_{c,N}^{(124)}(z\to z_1)$. Therefore
\begin{align}
\phi_{c,N}(z\to z_1)    =&  - 4 \eta_H \log |z-z_1| -\log\lambda +2\pi i \(N + \frac{1}{2}\) + 2\log(1- 2\eta_H) \nonumber\\ 
 &  + (1- 2\eta_H) \log \frac{ | z_{23}  |  | z_{24}  |}{ | z_{12} |^2 |z_{13} ||z_{14} | }  -\frac{4}{(1-2 \eta_H)}\eta_L -  \frac{2(1-\eta_H)}{z_{12}} (z -z_1)  \nonumber\\
&  + \frac{ \eta_L}{ \eta_{H} } \frac{1}{z_{12}}  \( \frac{z_{23}}{z_{13} } + \frac{z_{24}}{ z_{14}} \)  (z -z_1) +\dots \label{eq:asym1}
\end{align}
Recall that the only contributions from  $\phi_{c,N}^{(123)}$ and  $\phi_{c,N}^{(124)}$ that are effectively summed over for computing  $\phi_{c,N}$ are those linear in $\eta_L$. Also 
notice that in the second line we have just rewritten in a symmetric fashion the term $2(1- 2\eta_H) \log(  | z_{23}  |  /  |z_{13} | | z_{12} | )  =2(1- 2\eta_H) \log ( | z_{24}  |/ |z_{14} | | z_{12} |)  $ (where the equality follows from the circle of Apollonius condition).
One gets the asymptotic expansion around the heavy insertion at $z_2$ by replacing $z_2 \to z_1$ in \eqref{eq:asym1}. 
Using this expression we can derive the on-shell action up  to $\eta_H$-independent terms through  \eqref{eq:sigma12}.   It is easy to check that this is consistent with   \eqref{eq:Sconst}.

In the next section we will work out the functional dependence of the accessory parameters appearing in \eqref{eq:phiasymptotic} and in the relation $\eqref{eq:intC}$. At linear order in $\eta_L$, these can be read out from \eqref{eq:asym1} and \eqref{eq:asym3}-\eqref{eq:asym4}.  Here we quote the result for later reference and comparison
\begin{align}
c_1&= \frac{2\eta_H(1-\eta_H)}{z_{12}}-  \frac{\eta_L}{z_{12} } \( \frac{z_{23}}{ z_{13}} + \frac{z_{24}}{z_{14}} \) \label{eq:c1} \\ 
c_2&= \frac{2\eta_H(1-\eta_H)}{z_{21}}-    \frac{\eta_L}{z_{21} } \( \frac{z_{13}}{ z_{23}} + \frac{z_{14}}{z_{24}} \) \\ 
c_3&= -\eta_L \( \frac{1}{z_{13}}+ \frac{1 } { z_{23}} \) \\
c_4&=- \eta_L\( \frac{1}{z_{14}}+ \frac{1 } { z_{24}} \)\,. \label{eq:c4}
\end{align}

%%%%%%%%%%%%%%%%%%%%%%%%%%%%%%%%%%%%%%%%%%%%%%%%%%%%%%%%%%%%%%%%%%%%%%%
%%%%%%%%%%%%%%%%%%%%%%%%%%%%%%%%%%%%%%%%%%%%%%%%%%%%%%%%%%%%%%%%%%%%%%%%
\section{Monodromy method}
\label{sec:mono}

In this section we extend to the whole plane the saddles we have determined in Sec.~\ref{sec:linearized} on the circle of Apollonius. We use a standard approach, which consists in translating the problem of computing the semiclassical on-shell action with heavy insertions into a monodromy problem for the solutions of two ordinary second order differential equations. 

In the case $\l>0$ and for four insertions satisfying $\sum_i \eta_i >1$ with $\eta_i$ real, the monodromy method gives  the unique real, single-valued solution to the Liouville equation, and thus the unique real and single-valued saddlepoint. The range of conformal weights we are considering instead satisfies $\sum_i \eta_i <1$ with $\eta_i$ real. In this case, a unique real, single-valued solution would exist if and only if $\l <0$. Indeed in the previous section we have found that the solutions that can contribute to the saddle-point evaluation of the HHLL correlator  are single-valued but complex. In fact, on the circle  of Apollonius the Liouville field $\phi_{c,N}$ has constant imaginary part given by $2\pi i \(N + 1/2\)$, as was the case for the complex saddles discussed by  \cite{Harlow:2011ny} in the same regime of conformal dimensions. 

Notice however that the net effect of such constant imaginary part in the Liouville equation with heavy insertions \eqref{eq:Lsphere}
\be \label{eq:liouvilleeqbis}
\del_z \del_{\bar z} \phi_c = 2 \lambda e^{\phi_c} - 2\pi \sum_{i=1}^4 \eta_i \delta^{(2)}(z -z_i)
\ee
is to simply flip the sign in front of the exponential term, or equivalently  to replace $\lambda \to \tilde \lambda = -\lambda$. 
This means that, under the assumption that the imaginary part is of the form  $2\pi i \(N + 1/2\)$  in the full plane,  we can effectively solve for the real part of this set of complex solutions by using methods that are available in the literature for finding the unique real, single-valued solution with $\tilde \lambda <0$. 

Using this strategy, the solutions of the Liouville equation \eqref{eq:Lsphere} for the real part of the Liouville field with four heavy insertions satisfying $\sum_i \eta_i <1$  can be obtained as (see e.g. \cite{ZZ})
\be \label{eq:realphi}
\textrm{Re} \, \phi_c (z, \bar z) = - 2 \log   \(\psi_1 (z) \tilde{\psi}_1(\bar z)+ \psi_2 (z) \tilde{\psi}_2(\bar z)\)  - \log \lambda
\ee
where $ \psi_{1,2}(z), \tilde{\psi}_{1,2}(\bar z)$ are independent solutions with unit Wronskian of the system of equations
\bea
\psi''(z) +T(z) \psi(z) &=& 0 \label{eq:monhol} \\
\tilde \psi''(\bar z) +\tilde T(\bar z) \tilde \psi(\bar z) &=& 0\,. \label{eq:mopnanti}
\eea
Here $T(z),\tilde T(\bar z)$ are meromorphic and anti-meromorphic functions related to the  stress tensor. They are determined by asking that the solutions have the right singular behaviour  \eqref{eq:singular} at the insertion points and satisfy  the regularity condition \eqref{eq:BC} at infinity. Moreover, one requires the absence of further singularities, as these would not have a physical interpretation.

For two and three heavy operators insertions, these requirements completely fix the form of $T(z),\tilde T(\bar z)$. For more than three heavy insertions the conditions above imply 
\be
T(z) =\sum_{i }  \frac{\epsilon_i}{(z-z_i)^2} - \frac{c_i}{z-z_i}
\ee
with
\be
\eps_i \equiv  \frac{6h_i}{c} = \eta_i (1-\eta_i)  + O\( 1/ c\)  \,, \label{eq:epsidef}
\ee
subject to the constraints
\be \label{eq:constr}
\sum_i c_i=0\,, \quad \sum_i \left( c_i z_i - \epsilon_i \right) =0\,, \quad \sum_i \left( c_i z_i^2 - 2\epsilon_i z_i \right) =0 \, .
\ee
Similar equations hold for $\tilde T(\bar z)$. 
The $c_i$ are the accessory parameters we introduced in Sec.~\ref{sec:review} and are related via the Polyakov relation \eqref{eq:intC} to the derivative of the on-shell action. 
They depend on the coordinates of the singular points, as well as on the parameters $\eta_i$: $c_i = c_i(\{z_i, \bar z_i, \eta_i\})$. In particular the dependence is not holomorphic.

In the specific case of four heavy insertions, the constraints \eqref{eq:constr} fix all but one of the accessory parameters. 
The explicit constraints in terms of $c_4$ read
\begin{align}
c_{1} &=- \frac{ z_{24}z_{34} c_4 -2 [z_{13} \eta_H(1-\eta_H)   - z_{24} \eta_L] }{z_{12} z_{13}} \\
c_{2} &=-\frac{ z_{14}z_{34}c_4 - 2  [ z_{23}\eta_H(1-\eta_H) -z_{14} \eta_L ] }{z_{12} z_{23}} \\
c_{3}&=-\frac{ z_{14}z_{24}c_4 + 2 \eta_L (z_{13}+z_{24})}{z_{13} z_{23}}\ , 
\end{align} 
and are indeed satisfied by the accessory parameters \eqref{eq:c1}-\eqref{eq:c4} we computed in the previous section.

The general solutions of   the differential problem  \eqref{eq:monhol}-\eqref{eq:mopnanti} have $SL(2, {\mathbb C})$ monodromy matrices associated to closed paths, which depend on the free accessory parameter $c_4$.  For 
real $\eta_i$ with  $\sum_i \eta_i < 1$
where we solve an auxiliary Liouville equation with
$\tilde\lambda<0$, 
it is possible to construct  the unique real and single-valued solution to the Liouville equation through \eqref{eq:realphi} by finding a basis of solutions  $\{\psi_1(z), \psi_2(z)\}, \{\tilde \psi_1(\bar z), \tilde \psi_2(\bar z)\}$   that has $SU(2)$ monodromy around all cycles. (The more familiar condition that applies in the case of 
$\sum_i \eta_i > 1$ is instead that of $SU(1,1)$ monodromy.) 

The plan of the computations presented in this section is as follows. The solutions of the differential equations \eqref{eq:monhol}-\eqref{eq:mopnanti} up to linear order in $\eta_L$ were worked out in \cite{Fitzpatrick:2014vua}, which computed the semiclassical HHLL conformal blocks of perturbatively light exchanged operators. We review them and their monodromy transformations below for canonical insertions points, $z_1 = 1, z_2 = \infty, z_3 = 0, z_4 =x$. We then perform a change of basis such that the monodromy matrices around all cycles are unitary. This fixes unambiguously the accessory parameter $c_x = c_4$ at linear order in $\eta_L$, and thus also the functional dependence of the complex saddles that can contribute to the four-point function through 
\be
\frac{\del \tilde S}{\del x} =c_x \,, \qquad  \frac{\del \tilde S}{\del \bar x} = \bar c_x\,. 
\ee
%

%%%%%%%%%%%%%%%%%%%%%%%%%%%%%%%
\subsection{Solutions and monodromy matrices}  

For canonical insertions points  $z_1 = 1, z_2 = \infty, z_3 = 0, z_4 =x$ we have 
\be
T(z) = \frac{\epsilon_H}{(z-1)^2} + \epsilon_L\left( \frac{1}{z^2} + \frac{1}{(z-x)^2} + \frac{2}{z(1-z)} \right) - \frac{c_x \, x (1-x)}{z(z-x)(1-z)}\, .
\ee
We are not able to solve the second order differential equation for generic $\epsilon_{i}$, but \cite{Fitzpatrick:2014vua} solved it for $\epsilon_1 = \epsilon_2 = \epsilon_H$, $\epsilon_3 = \epsilon_4= \epsilon_L$ to all orders in $\epsilon_H$ and to linear order in $\epsilon_L$.   
To this order we have that  $\epsilon_L \approx \eta_L$,  and a basis of solutions for \eqref{eq:monhol} is given by \cite{Fitzpatrick:2014vua} %
\bea
\psi_1(z) &=& (1-z)^{\frac{1+ \tilde\alpha}{2}} + \eta_L \Big[ (1-z)^{\frac{1+\tilde\alpha}{2}} \frac{\left(\frac{c_x}{\eta_L} (1-x)+1 \right)\log \frac{z}{z-x} + \frac{(x-2)z+x}{z(z-x)}}{\tilde\alpha}  \label{psi1}\\ 
&& + (1-z)^{\frac{1- \tilde\alpha}{2}} \int dz \frac{(1-z)^{\tilde\alpha} \left( \frac{c_x(x-1)x z (x-z)}{\eta_L} - x^2 (z+1)+ 2 x z (z+1)-2z^2 \right)}{z^2 \tilde\alpha(x-z)^2}\Big] \nonumber \, \  \\
\psi_2(z) &=& (1-z)^{\frac{1- \tilde\alpha}{2}} - \eta_L \Big[  (1-z)^{\frac{1- \tilde\alpha}{2}} \frac{\left(\frac{c_x}{\eta_L} (1-x)+1 \right)\log \frac{z}{z-x} + \frac{(x-2)z+x}{z(z-x)}}{\tilde\alpha} \label{psi2} \\
&&  +(1-z)^{\frac{1+\tilde\alpha}{2}} \int dz \frac{(1-z)^{-\tilde\alpha} \left( \frac{c_x(x-1)x z (x-z)}{\eta_L} - x^2 (z+1)+ 2 x z (z+1)-2z^2 \right)}{z^2 \tilde\alpha(x-z)^2}  \Big]  \nonumber 
\eea
where 
\be
\tilde\alpha \equiv \sqrt{1-4 \epsilon_H} = 1- 2\eta_H + O\left(1/c \right)
\ee 
and  $\psi_2(z)[\ta] = \psi_1(z)[-\ta] $. 
The integral above can also be performed explicitly in terms of hypergeometric functions. 

In order to fix the accessory parameter in terms of the monodromy properties, we first need to know how  this basis of solutions transforms.  To fix the notation, taking $z$ in a closed loop  $\gamma$ around a point $z_i$, 
 the solution will have monodromy $M_{\gamma}$
\be
\( \begin{array}{c} \psi_1(z) \\  \psi_2(z) \end{array}\) \to M_{\gamma} \( \begin{array}{c} \psi_1(z) \\  \psi_2(z) \end{array}\) \, , 
 \ee
where the matrix $M_{\gamma} \in SL(2,{\mathbb C})$. The details of the computation are reported in Appendix~\ref{app:monocomp}, and here we just summarize the resulting monodromies working under the assumption that $\tilde\alpha \in \mathbb{R}$.

The transformations of the solutions upon taking them in a closed loop around $z=0$ and around $z=x$ can be immediately evaluated by noticing that the only non-trivial monodromies are those given by the $\log$ and the integrals appearing in \eqref{psi1} and \eqref{psi2}, which, as in \cite{Fitzpatrick:2014vua}, can be evaluated from the residues. The monodromy transformations at linear order in $\eta_L$ are given by  
\bea
 M_{\gamma_0} \!\!&=&\!\! \Id+ \frac{2\pi i}{\tilde\alpha}
 \left(\begin{array}{cc} \cx(1-x) + \eta_L & -\cx(1-x) - \eta_L (1-\tilde\alpha) \\
\cx(1-x) + \eta_L (1+\tilde\alpha)&  -\cx(1-x) - \eta_L\end{array}\right) \label{eq:M0} \\
{} \nonumber \\{} \nonumber \\
 M_{\gamma_x} \!\!&=&\!\! \Id + \frac{2\pi i}{\tilde\alpha}
 \left(\begin{array}{cc} -\cx(1-x) - \eta_L & (1-x)^{\tilde\alpha} \(  \cx(1-x) + \eta_L (1+\tilde\alpha) \) \\
 (1-x)^{-\tilde\alpha} \( -\cx(1-x) - \eta_L (1-\tilde\alpha) \)   &  \cx(1-x) +\eta_L \end{array}\right) \, . \label{eq:Mx}\nonumber \\
\eea
The monodromy around $z=1$ requires a little more work and can be computed by performing the integrals in \eqref{psi1} and \eqref{psi2}  and expressing them in terms of appropriate combinations of hypergeometric functions (see Appendix~\ref{app:monocomp}). The resulting expression at linear order in $\eta_L$ is 
\begin{align}
& M_{\gamma_1} = - \left(\begin{array}{cc} e^{\tilde\alpha\pi i} & 0 \\0 & e^{-\tilde\alpha \pi i}\end{array}\right)  \label{eq:M1} \\
&+ \frac{2\pi i}{\tilde\alpha}  
 \left(\begin{array}{cc} 0 &  \!\!\!\!\!\!\!\!\! \!\!\!\!\!\!\!\!\!   \!\!\!\!\!\!\!\!\! \!\!\!\!\!\!\!\!\! e^{ \tilde\alpha \pi i}\left[  - \left(\cx (1-x)+\eta_L \right) \(1-(1-x)^{\tilde\alpha}\) + \eta_L \tilde\alpha\right] \\ 
e^{- \tilde\alpha \pi i}\left[ \left( \cx(1-x)+ \eta_L \right) \(1-(1-x)^{-\tilde\alpha} \)+ \eta_L \tilde\alpha\right]  &   \!\!\!\!\!\!\!\!\! \!\!\!\!\!\!\!\!\!  \!\!\!\!\!\!\!\!\! \!\!\!\!\!\!\!\!\! 0
\end{array}\right) \, . \nonumber 
\end{align}
%

%%%%%%%%%%%%%%%%%%%%%%%%%%%%%%%%%%%%%%%%%%%%%%%%%
\subsection{Accessory parameter for $SU(2)$ monodromy}

The unique real and single-valued solution to the Liouville equation with $\tilde\lambda < 0$ is  not directly associated to   $\psi_1$ and $\psi_2$ above, but to a basis of solutions to \eqref{eq:monhol} with $SU(2)$ monodromy about every insertion. We therefore consider a change of basis 
\be \label{eq:su2B}
B \in SL(2, {\mathbb C}) \qquad \mbox{ such that } \qquad  N_{\gamma} \equiv  BM_{\gamma} B^{-1} \in SU(2) \qquad \forall \gamma\, 
\ee
for an appropriate choice of the accessory parameter. 
While working in the linearized approximation, a convenient  way of imposing \eqref{eq:su2B} is just to rewrite the unitarity condition on $N_{\gamma}$ as 
\be \label{eq:JJ}
\tilde J M_\gamma = (M_\gamma^{-1})^\dagger \tilde J \\
\ee
with
\be
 \tilde J =  \left(\begin{array}{cc}
a & b \\
\bar{b}& d
\end{array}\right)\, 
\equiv  B^\dagger  B \, , 
\ee 
which implies  $a,d \in {\mathbb R}$  and  $\det \tilde J = 1$.

Working perturbatively in $\eta_L$ up to linear order, we split $M_\gamma = M_\gamma^{(0)} +\eta_L \delta M_\gamma$ and similarly for $B$ and $\tilde J$, and solve at each order for $\tilde J$. At zeroth order we have the trivial monodromies
\be
M^{(0)}_{\gamma 0} = M^{(0)}_{\gamma x} = \Id, 
\ee
which make \eqref{eq:JJ} trivially satisfied, while  
\be
M^{(0)}_{\gamma_1} = \left(\begin{array}{cc} e^{(1+\tilde\alpha)\pi i} & 0 \\0 & e^{(1-\tilde\alpha) \pi i}\end{array}\right) 
\ee
imposes the zeroth order constraint $b^{(0)} =0$, and therefore $a^{(0)} d^{(0)} =1$. That is,  
\be \label{eq:Jtilde0}
\tilde J^{(0)} = 
 \left(\begin{array}{cc}
a^{(0)} & 0 \\
0 & \frac{1}{a^{(0)} }
\end{array}\right) \, ,
\ee
with $a^{(0)}$ undetermined at this order. 

At linear order in $\eta_L$ equation \eqref{eq:JJ} reads 
\be   \label{eq:JJ1}
\tilde J^{(0)}  \delta M_\gamma - \delta(M_\gamma^{-1})^\dagger \tilde J^{(0)} = M_\gamma^{(0)} \delta \tilde J - \delta \tilde J M_\gamma^{(0)} \, ,
\ee
where we have used the fact that $M_\gamma^{(0)} = (M_\gamma^{(0)-1})^\dagger $.  
The linear parts in $\eta_L$  of the monodromy matrices are read from  \eqref{eq:M0}, \eqref{eq:Mx} and \eqref{eq:M1}.
For the monodromy around $0$ and  $x$, $M_\gamma^{(0)} = \Id$ so that \eqref{eq:JJ} further simplifies to 
\be
\tilde J^{(0)}  \delta M_\gamma = \delta(M_\gamma^{-1})^\dagger \tilde J^{(0)}\,.
\ee
Imposing these matrix equations, one gets the reality condition 
\be
\cx(1-x) =\bc_x(1-\bx) \, .
\ee
Moreover, using the above constraint, the equations for the insertions at $0$ and $x$ imply also
\bea
(a^{(0)})^2 &=& -\frac{ \frac{ \cx}{\eta_L} (1-x)+1+\tilde\alpha}{ \frac{\cx}{\eta_L} (1-x)+1-\tilde\alpha} \\
(a^{(0)})^2 &=& - \frac{1}{|1-x|^{2\tilde\alpha}} \frac{\frac{ \cx}{\eta_L} (1-x)+1-\tilde\alpha}{\frac{\cx}{\eta_L} (1-x)+1+\tilde\alpha}\,,
\eea
which have solutions
\bea
\frac{\cx}{\eta_L} &=& \frac{\mp (1+\tilde\alpha)|1-x|^{\tilde\alpha} -1 +\tilde\alpha}{(1-x)(1 \pm |1-x|^{\tilde\alpha})}\\
(a^{(0)})^2 &=& \pm \frac{1}{|1-x|^{\tilde\alpha}} \, . \label{eq:a0squared}
\eea
Remembering that  $a^{(0)}$ must be real  the upper sign in \eqref{eq:a0squared} is selected and this fixes
\be \label{eq:cSU(2)}
\frac{\cx}{\eta_L} = \frac{- (1+\tilde\alpha)|1-x|^{\tilde\alpha} -1 +\tilde\alpha}{(1-x)(1+ |1-x|^{\tilde\alpha})}\,. 
\ee
This indeed coincides with the accessory parameter \eqref{eq:c4} which we worked out in the previous section on the Apollonius circle  for canonical insertion points,  $|1-x|=1$.

%%%%%%%%%%%%%%%%%%%%%%%%%%%%%%%%%%%%%%%%%%%%%%%%%
\subsection{Functional dependence of the saddles}

Having  determined  in $\eqref{eq:cSU(2)}$ the accessory parameter associated to the insertion at $x$, one can integrate its expression with respect to  $x$  and obtain the semiclassical action up to a function of $\bx$,  
\be
\tilde S(x, \bar x) = 4 \eta_L \log \(1+ |1-x|^{\tilde\alpha}\) + \eta_L (1-\tilde\alpha) \log (x-1) + K(\bar x)\,. 
\ee
Repeating the analysis for the anti-meromorphic counterpart with $\tilde T(\bar z)$, one similarly finds 
\be
\tilde S(x, \bar x) = 4 \eta_L \log \( 1+ |1-x|^{\tilde\alpha}\) + \eta_L (1-\tilde\alpha) \log (\bar x-1) + \tilde K( x)\,.
\ee
Matching the two we obtain the on-shell action up to constant terms,
\be \label{eq:SSU(2)}
\tilde S(x, \bar x) = 4 \eta_L \log \( 1+ |1-x|^{1-2\eta_H}\) + 4 \eta_L \eta_H \log |1-x| + \textrm{const} \,, 
\ee
where we substituted back  $\tilde \a = 1- 2\eta_H$. 

On the unit circle this reduces to \eqref{eq:Sconst}, which fixes the constant term.  
We therefore arrive at the explicit form of possible contributing saddlepoints:
\begin{align} 
\tilde{S}{_N}(x, \bar x)  =& - (1- 2\eta_H- 2\eta_L) \log \lambda + 2\pi i \(N+\frac{1}{2}\) ( 1-2\eta_H - 2\eta_L) - 2 (1-2\eta_H) \nonumber \\
& + 2 (1-2\eta_H - 2\eta_L) \log(1-2\eta_H)  +4\eta_H(1-\eta_H)\log |z_\infty|  \nonumber \\ 
&+  4 \eta_L \log \( 1+ |1-x|^{1-2\eta_H}\) + 4 \eta_L \eta_H \log |1-x| + \dots \label{eq:superpmono}
\end{align} 
where $N$ is any integer. These are all complex saddles consistent with the assumption that the imaginary part of the Liouville field is of the form $2\pi i \(N + 1/2\)$. 

The computations we have performed so far have allowed us to fix the form of the possible saddlepoint contributions, but do not directly determine which set of saddles should actually be summed to compute the HHLL correlator. For this purpose, in the next section we analyze the conformal block decomposition of the correlator.
  
%%%%%%%%%%%%%%%%%%%%%%%%%%%%%%%%%%%%%%%%%%%%%%%%%%%%%%%%%%%%%%%%%%%%%%%%
%%%%%%%%%%%%%%%%%%%%%%%%%%%%%%%%%%%%%%%%%%%%%%%%%%%%%%%%%%%%%%%%%%%%%%%%
\section{Conformal block expansion}
\label{sec:DOZZ}

A standard way  of  decomposing a four-point function in a  CFT is  via a conformal block expansion. In Liouville theory such a decomposition is subtle due to the continuous spectrum, and to the fact that some care is needed when defining normalizable states and a complete set thereof.
Concretely, a four-point function 
\be \label{eq:G}
G(x,\bx)  \equiv \lim_{ z_\infty \to \infty} |z_\infty|^{ 4 h_2 } \langle  V_{\a_1}(1)   V_{\a_2}(z_\infty, \bz_\infty)  V_{\a_3}(0) V_{\a_4}(x,\bx)  \rangle  \, ,
\ee
for real momenta $\a_i$ less than $Q/2$  and such that  $\alpha_1+\alpha_2> Q/2$ and $\alpha_3+\alpha_4  > Q/2$,  can be expanded in conformal blocks as  \cite{Seiberg:1990eb,Zamolodchikov:1995aa}  
\begin{equation}
G(x,\bar{x}) = \frac 1 2 \int_{-\infty}^{\infty} \frac{dP}{2\pi} \, C\(\alpha_1,\alpha_2,\frac Q 2- iP\)C\(\alpha_3,\alpha_4,\frac Q 2 + iP\)\mathcal{F}(h_i,h_P,x) 
{\mathcal{F}}(h_i,h_P,\bx) \, ,   \label{eq:ConfBlock}
\end{equation}
with $\mathcal{F}$ denoting the Virasoro conformal block, $h_i$ the conformal weights of the external operators and $h_P= Q^2/4+ P^2$ the weights of the primaries exchanged in the intermediate channel.  $Q$ is related to the central charge via \eqref{eq:Qdef}, but $P$ is arbitrary.  In this regime of momenta, external operators define normalizable states. The expansion is given in terms of intermediate normalizable states, which are heavy, since $h_P \ge c/ 24$ in the semiclassical limit.

The functions $C(\a_1,\a_2,\a_3)$ 
are the three-point function coefficients, which are given by the DOZZ formula \cite{Dorn:1994xn,Zamolodchikov:1995aa}:
\begin{align} \label{eq:FusCoef}
C(\alpha_1,\alpha_2,\alpha_3) &= \[ \pi \mu \gamma (b^2) b^{2-2b^2}\]^{(Q-\sum_i \alpha_i)/b} \\
& \times\frac{ \Upsilon_0 \Upsilon_b(2\alpha_1) \Upsilon_b(2\alpha_2) \Upsilon_b(2\alpha_3)}
{\Upsilon_b (\alpha_1+\alpha_2+\alpha_3-Q)\Upsilon_b (\alpha_1+\alpha_2-\alpha_3) \Upsilon_b (\alpha_1-\alpha_2+\alpha_3) \Upsilon_b (-\alpha_1+\alpha_2+\alpha_3)  }\,. \nonumber 
\end{align}
Here 
$\Upsilon_b$ is defined for $b>0$ as
\be
\log \Upsilon_b (x) = \int_{0}^{\infty} ~ \frac{dt}{t} \[ \(\frac{Q}{2}- x\)^2 e^{-t }- \frac{\sinh^2\[   (Q/2-x) \frac{t}{2}\]}{\sinh\( \frac{t b}{2} \) \sinh\(\frac{t}{2b} \) }\]    
\ee  
when $0< \Re(x)< Q$, and 
\be
\Upsilon_0 \equiv \frac{d \Upsilon_b (x)}{dx} \Big|_{x=0}\,.
\ee 
We will list some of the properties of the $ \Upsilon_b$ function as we use them in the following, and refer to Appendix A of \cite{Harlow:2011ny} for a detailed review. 

We are interested in the semiclassical limit $b \to 0$, where we consider heavy insertions with real momenta $\alpha_1=\alpha_2= \a_H= \eta_H / b$ and $\alpha_3=\alpha_4= \a_L =\eta_L /b$.  We also assume $1/4<\eta_H < 1/2$, such that  $\alpha_1+\alpha_2  > Q/2$ satisfies  the normalizability condition, and $\eta_L \ll1$, for which instead  $\alpha_3+\alpha_4<Q/2$. Nevertheless, it is possible to define such four-point functions starting from \eqref{eq:ConfBlock} by analytic continuation in $\eta_L$ \cite{Zamolodchikov:1995aa}.  The key point in such a continuation concerns poles in the integrand of the $P$ integral.   In the original regime of applicability of  \eqref{eq:ConfBlock} ($1/4<\eta_L < 1/2$), the function $C$  has poles in the complex plane that do not lie on the contour of integration, namely the  real $P$ axis.  As we continue $\eta_L$ outside this regime, some of the poles cross the contour of integration. Their residues contribute discrete terms in the conformal block expansion, corresponding to additional lighter non-normalizable exchanges. That is,  upon analytic continuation  we have to evaluate schematically  
\begin{align} \label{eq:blocksch}
G(x,\bar{x})= &~ \frac{i}{2}\sum_{\substack{ {\rm crossing} \\ {\rm poles} }} C\(\alpha_1,\alpha_2,\frac Q 2 - iP\)  {\rm Res~}  C\(\alpha_3,\alpha_4,\frac Q 2+iP\)\mathcal{F}(h_i,h_P,x) 
{\mathcal{F}}(h_i,h_P,\bx)   \\
&+\frac 1 2 \int_{-\infty}^{\infty} \frac{dP}{2\pi} \, C\(\alpha_1,\alpha_2,\frac Q 2-iP\)C\(\alpha_3,\alpha_4,\frac Q 2+iP\)\mathcal{F}(h_i,h_P,x) 
{\mathcal{F}}(h_i,h_P,\bx) \, . \nonumber
 \end{align}
%
 
 %%%%%%%%%%%%%%%%%%%%%%%%%%%%%%%%%%%%%%%%%%%%%%% 
 \subsection{Discrete contributions to the block decomposition} \label{sub:discrete}
 
 In order to understand which additional discrete intermediate states appear in the expansion  of the HHLL correlator,  consider  the pole structure of 
\be \label{eq:Cpoles}
C\(\alpha_L,\alpha_L,\frac Q 2 + iP\) \propto \frac{\Upsilon_0 \Upsilon^2_b(2\alpha_L)\Upsilon_b(Q + 2 iP))}{ \Upsilon_b(2\alpha_L - Q/2 + iP)\Upsilon_b(2\alpha_L - Q/2- iP )\Upsilon^2_b(Q/2 + iP )}\,.
\ee
Poles in this expression come from simple zeros of the  $\Upsilon_b(x)$. As can be shown using $\Upsilon_b(x)$ recursion relations,  these occur at $x = - m b- n/b$ and  $x = (\tilde m +1) b + (\tilde n+1)/b$, with $m,\tilde m, n, \tilde n$ non-negative integers \cite{Zamolodchikov:1995aa,Harlow:2011ny}.   In particular, poles which can cross the real $P$ axis upon analytic continuation in  $\eta_L$ originate from zeros of   $\Upsilon_b(2\alpha_L - Q/2 + iP)$ and $\Upsilon_b(2\alpha_L - Q/2- iP)$ in the denominator.
These are located on the imaginary $P$-axis at:
\bea \label{eq:poles}
\pm \, i P &=&  - 2 \a_L + \frac Q 2 - m b - \frac n b\\
\pm \, i P &=& -  2 \a_L + \frac Q 2  + (\tilde m +1) b +\frac{\tilde n+1}{b} \,.
\eea
In the semiclassical limit $b\to 0$ we effectively have infinite ``towers''  of $b$-spaced poles, labeled by $m$ and $\tilde m$, around  
\bea
\pm \, i P & \approx &  -2 \a_L-  \frac{ 2n-1}{2 b} \\
\pm \, i P &\approx & -  2 \a_L +\frac{2 \tilde  n+3}{2 b}
\eea
for each non-negative integer $n$ and $\tilde n$.

\begin{figure}[th]
\centering
\includegraphics[width=0.5\textwidth]{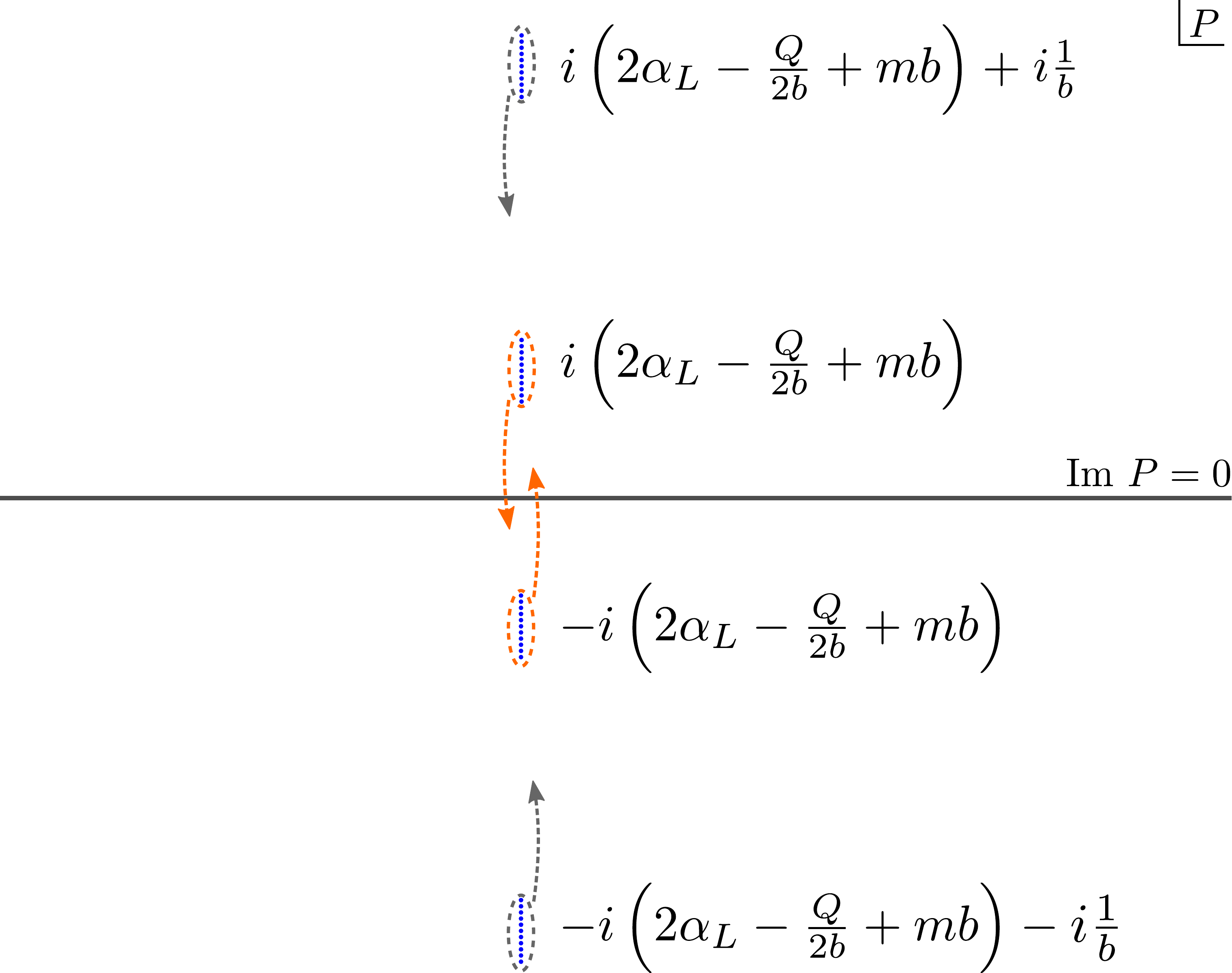}
\caption{ Representation of the poles \eqref{eq:poles} of the DOZZ coefficients in the complex $P$-plane.  The poles come in towers, represented by the dots inside the dashed ellipses, where  elements  are labeled by  non-negative integers $m$ and spaced by $b$.     The arrows indicate the direction of the analytic continuation as one continues $\alpha_L$ from $\alpha_L >  Q/4 \approx 1/4 b$ to the regime $\alpha_L <  Q/4$.  In the  semiclassical limit $b\to 0$, only the two towers of poles  with $n=0$ in \eqref{eq:poles}  cross the real $P$-axis.}\label{fig:PoleStructure}
\end{figure}

As we analytically continue $\eta_L$ to the regime $\eta_L \ll 1$, the $n=0$ poles (for all $m=0, \dots, \infty$ in the $b \to 0$ limit) in \eqref{eq:poles} change sign and cross the real $P$-axis  (see Fig.~\ref{fig:PoleStructure}), and therefore one needs to deform the contour of integration in \eqref{eq:ConfBlock}. 
 Equivalently, the analytic continuation in $\eta_L$ adds to the expansion of the four-point function discrete terms given by the residues of the poles at
\be \label{eq:polesRes}
i P = \mp \(2\alpha_L - \frac{Q}{2}+ m b\)\,, \quad \quad m =0 , \dots, \infty\,.
\ee
Both sets of  extra contributions to the block decomposition correspond to internal operators of weight $h_{P_m} = \a_{P_m} (Q-\a_{P_m})$, with $\a_{P_m} \equiv    2\a_L + m b$.\footnote{ Remember exponential operators with Liouville momenta $\a$ and $Q-\a$   correspond to the same operator with weight $h_\a$. 
}   
It follows  
that in the semiclassical limit
\be\label{eq:hPmv0}
h_{P_m} = \frac{2 \eta_L(1-2\eta_L)}{b^2} + m + 2\eta_L(1-2m) + O(b^2) \,.
\ee

The first term  is $O(c)$ and the second term is $O(1)$, and thus suppressed by $1/c$.
However, since we have an infinite number of contributions, we need to take explicitly into account the effect of order one terms in the conformal weight  \eqref{eq:hPmv0} to understand how they  affect the leading semiclassical result. 
For consistency with the linearized results of the previous sections and with the order at which the semiclassical HHLL Virasoro conformal blocks ${\cal F}$ are known analytically \cite{Fitzpatrick:2015zha}, we will neglect terms of $O(\eta_L^2/b^2)$ and $O(\eta_L)$, while keeping  $O(\eta_L/b^2)$ and $O(1)$ in the above formula. 
The statement that we are keeping order one terms while neglecting terms $\sim h_L^2/c$ thus implies the assumption  $h_L^2/c \sim \eta_L^2/b^2 \lesssim1$.\footnote{This same assumption was also made and discussed in \cite{Fitzpatrick:2014vua} when computing HHLL semiclassical Virasoro blocks.} 
In the following we will thus approximate
\be \label{eq:hPm}
h_{P_m}\approx \frac{2 \eta_L}{b^2} + m \,  , 
\ee
and work out the conformal block expansion consistently.  

We now evaluate explicitly the contributions of the two sets of poles \eqref{eq:polesRes}, corresponding to the exchange of operators with weight  $h_{P_m} $.
For the   upper choice of sign in \eqref{eq:polesRes},  $Q/2 -  i P = \a_{P_m}= 2\a_L + m b$ and  we have 
\begin{align}
C \(\alpha_H,\alpha_H, \frac Q 2-iP \)  = &\[ \pi \mu \gamma(b^2) b^{2-2b^2}\]^{1-m + \frac{1-2\eta_H-2\eta_L}{b^2}}    \\
&\times \frac{ \Upsilon_0 \Upsilon_b^2 (2\alpha_H) \Upsilon_b (4\alpha_L +2 m b)}{ \Upsilon_b (2\alpha_H + 2\alpha_L +m b -Q) \Upsilon_b (2\alpha_H - 2\alpha_L -m b) \Upsilon_b^2 (2\alpha_L +m b)} \nonumber  \ , 
\end{align}
and
\begin{align}
2\pi i \, {\rm Res} \, C\(\alpha_L,\alpha_L,\frac Q 2 + i P\) = &2\pi \[ \pi \mu \gamma(b^2) b^{2-2b^2}\]^{m} b^{m[1+b^2 (1+m)]} \(\prod_{j=1}^{m} \gamma(-jb^2)\)  \nonumber  \\
&\times \frac{ \Upsilon_b^2 (2\alpha_L) \Upsilon_b (2Q-4\alpha_L -2mb)}{ \Upsilon_b (4\alpha_L  +mb -Q)  \Upsilon_b^2 (Q-2\alpha_L -mb)}\,.  \label{eq:resC}
\end{align}
Here we have used the recursion relation 
\be \label{eq:upsilonminusb}
\Upsilon_b (x- b) = \gamma(b x - b^2)^{-1} b^{2b x -1-2b^2} \Upsilon_b(x) 
\ee
and the fact that $\Upsilon_b (x)$ vanishes linearly
\be
\Upsilon_b (x) \approx x \Upsilon_0\, \quad \quad {\rm as} \quad x \to 0 \, .
\ee
The product $\prod_{j=1}^{m} \gamma(-jb^2)$ in \eqref{eq:resC} is understood to evaluate to $1$ for $m=0$. 

The poles associated to the lower choice of sign in  \eqref{eq:polesRes} have instead $Q/2 -  i P = Q -\a_{P_m} = Q - 2\a_L - m b$,  and  it is  easy to see that
\begin{align}
C(\alpha_H,  \alpha_H, 2\alpha_L + m b )& \textrm{Res }  C(\alpha_L, \alpha_L, Q- 2\a_L - m b) = \nonumber \\
&-  C(\alpha_H,  \alpha_H, Q- 2\a_L - m b) \textrm{Res }  C(\alpha_L, \alpha_L, 2\alpha_L + m b )\,,
\end{align}
which is equal to minus the contribution of the other set of poles in  \eqref{eq:polesRes}.
On the other hand the contours  have opposite orientations and therefore they lead to two identical contributions.

The block decomposition in our regime is thus obtained as 
\begin{align}
G(x,\bar{x}) &=   i \sum_{m=0}^\infty C(\alpha_H,  \alpha_H, 2\alpha_L + mb ) \textrm{Res }  C(\alpha_L, \alpha_L, Q- 2\a_L - mb) \mathcal{F}(h_i,h_{P_m},x)
{\mathcal{F}}(h_i,h_{P_m},\bx)  \\ 
&+ \frac{1}{2}\int_{-\infty}^{\infty} \frac{dP}{2\pi} \, C\(\alpha_H,  \alpha_H,  \frac{Q}{2}- i P \) C\(\alpha_L, \alpha_L, \frac{Q}{2} + i P\) \mathcal{F}(h_i,h_P,x)
{\mathcal{F}}(h_i,h_P,\bx) \,, \nonumber
\end{align}
with each discrete term evaluating  to
\begin{align}
&2\pi iC(\alpha_H,  \alpha_H, 2\alpha_L + m b ) \textrm{Res }  C(\alpha_L, \alpha_L, Q- 2\a_L - m b) = \nonumber\\
& 2\pi \left[ \pi \mu \gamma(b^2) b^{2-2b^2}\right]^{1 + \frac{1-2\eta_H-2\eta_L}{b^2}} b^{m[1+b^2 (1+m)]}    \(\prod_{j=1}^{m} \gamma(-jb^2)\)   \\
&\times\frac{\Upsilon_0 \Upsilon_b^2(2\alpha_H) \Upsilon_b^2 (2\alpha_L) \Upsilon_b(4\alpha_L+ 2m b) \Upsilon_b (2Q-4\alpha_L -2mb)}{ \Upsilon_b (2\alpha_H + 2\alpha_L +m b -Q) \Upsilon_b (2\alpha_H - 2\alpha_L -m b) \Upsilon_b^4 (2\alpha_L +m b) \Upsilon_b (4\alpha_L  +m b -Q)}\, ,   \nonumber
\end{align}
where we have used the property $\Upsilon_b(Q-x) = \Upsilon_b(x)$. 

%%%%%%%%%%%%%%%%%%%%%%%%%%%%%%%%%%%%%%%%%%%%%%%
 \subsection{Sum over the discrete terms} \label{sec:discrete}
 
To  explicitly  compute  the contribution of the discrete terms, we evaluate the series over $m$. For simplicity, in this section only, we set  $x = 1-w$ and $w = e^{2 i \phi_0}$. This is equivalent, via conformal transformation, to studying  the correlator with  canonical insertions in the complex $w$ plane with  light operators on the unit circle  and heavy operators at the origin and at infinity. In radial quantization this is directly related to a correlator of two   light insertions at equal time  in a heavy state. We generalize the analysis to arbitrary insertion points in Appendix \ref{app:doubletrace}, and state the general result at the end of this section.

For the range of conformal weights we are interested in, the semiclassical Virasoro blocks  read \cite{Fitzpatrick:2015zha}
\be
{\cal F}(h_i, h_{P_m},  1-w) = w^{({\tilde \a}-1)h_L} \left(\frac{1-w^{\tilde \a}}{{\tilde \a}} \right)^{h_{P_m} -2 h_L} \hf\left[ h_{P_m}, h_{P_m}, 2 h_{P_m}, 1-w^{\tilde \a} \right]\,,
\ee
where we recall that 
\be
{\tilde \a} =   1 - 2 \eta_H + O\( 1/ c\) \, .
\ee
On the unit circle we therefore have
\begin{align}
G(1-w,1-\bar{w}) \approx&  i \sum_{m=0}^\infty C(\alpha_H,  \alpha_H, 2\alpha_L + mb ) \textrm{Res }  C(\alpha_L, \alpha_L, Q- 2\a_L - mb)
\left(\frac{2 \sin (\tilde \a \phi_0)}{\tilde \a}\right)^{2m}\nonumber\\
& \times \hf\left[ h_{P_m}, h_{P_m}, 2 h_{P_m}, 1-e^{2 i \tilde \a \phi_0} \right] \hf\left[ h_{P_m}, h_{P_m}, 2 h_{P_m}, 1-e^{- 2 i \tilde \a \phi_0}  \right] \nonumber\\
\nonumber \\
&+ \,\mbox{integral over the continuous spectrum}\,. \label{eq:discrunit}
\end{align}
Defining $z \equiv 1-w^{\tilde \a} = 1-e^{2 i \tilde \a \phi_0} $, we can rewrite \eqref{eq:discrunit} as
\begin{align}
& G(1-w,1-\bar{w}) \approx  i \sum_{m=0}^\infty C(\alpha_H,  \alpha_H, 2\alpha_L + mb ) \textrm{Res }  C(\alpha_L, \alpha_L, Q- 2\a_L - mb)
 \tilde \a^{-2 m } \left( \frac{z^2}{z-1} \)^{m}\nonumber  \\
& \times \hf\left[ h_{P_m}, h_{P_m}, 2 h_{P_m}, z \right] \hf\left[ h_{P_m}, h_{P_m}, 2 h_{P_m}, \frac{z}{z-1}  \right]
+ \,  \mbox{continuous spectrum}\,.
\end{align}
The hypergeometric identity
\bea
\hf\left[ h_P, h_P, 2 h_P, z \right] \hf\left[ h_P, h_P, 2 h_P, \frac{z}{z-1}\right] = {}_{3}F_{2} \left( \left.\begin{array}{c} h_P,h_P,h_P  \\ 2h_P, h_P+\frac 1 2 \end{array}\right| \left.\begin{array}{c} \frac{z^2}{4(z-1)} \end{array} \right. \) 
\eea
allows us to write the sum over discrete exchanges as a series in powers of $y \equiv z^2 /(z-1)$, 
\begin{align}
&\sum_{m=0}^\infty \sum_{n=0}^\infty  C(\alpha_H,  \alpha_H, 2\alpha_L + mb ) \textrm{Res }  C(\alpha_L, \alpha_L, Q- 2\a_L - mb)  \frac{ \tilde \a^{-2m} 2^{-2n}y^{m+n} \left[ (h_{P_m})_n \right]^3 }{n! (2h_{P_m})_n (h_{P_m}+\frac 1 2)_n}  \nonumber \\
&  = \sum_{k=0}^\infty \tilde \a^{-2k} y^k \sum_{n=0}^k  A_{k-n} ~ \(\frac{ \tilde \a}{2}\)^{2n}  \frac{\left[ (h_{P_{k-n}})_n \right]^3 }{n! (2h_{P_{k-n}})_n (h_{P_{k-n}}+\frac 1 2)_n}\nonumber \\
&  = \sum_{k=0}^\infty \tilde \a^{-2k} y^k \beta_k,
\end{align}
where we introduced the notation  
\bea
A_j &\equiv& C (\alpha_H,  \alpha_H, 2\alpha_L + jb ) \textrm{Res }  C(\alpha_L, \alpha_L, Q- 2\a_L - jb)   \label{eq:AmDOZZ}\\
\beta_k &\equiv& \sum_{n=0}^k  A_{k-n} ~ ~ \(\frac{ \tilde \a}{2}\)^{2n}\frac{\left[ (h_{P_{k-n}})_n \right]^3 }{n! (2h_{P_{k-n}})_n (h_{P_{k-n}}+\frac 1 2)_n} \,.  \label{eq:betak}
\eea
Using the relation \eqref{eq:upsilonminusb} and 
\be
\Upsilon_b(x+b) = \gamma (bx) b^{1-2b x} \Upsilon_b(x)
\ee
we can work out a recursion formula for the $A_j$ coefficients 
\begin{align}
 \frac{A_{j+1}}{A_{j}}  =&
 \gamma[-b^2 (1 + j)] \gamma[-b^2 (1 + j) + 2 (\eta_H - \eta_L)]    \gamma[   2 + b^2(1 - j) - 4 \eta_L]   \nonumber \\
 &\times \gamma[ 1 - b^2 j - 2 \eta_L]^4\gamma[   2 b^2 j + 4 \eta_L]    \gamma[-1 + 2 b^2 j + 4 \eta_L]  \gamma[   b^2(1 + 2  j) + 4 \eta_L]  \nonumber \\
  &\times \gamma[-1 - b^2 (1 - 2 j) +     4 \eta_L] \gamma[b^2 (1 - j) + 2 (1 - \eta_H - \eta_L)] \,.
\end{align}
In the limit $b \to 0$ and up to terms of $O(\eta_L)$ and $O(\eta_L^2/b^2)$ this reduces to 
\be
\frac{A_{j+1}}{A_{j}} \approx  -\frac{(1- 2 \eta_H)^2 }{16 (1+j)} \( \frac {4 \eta_L}{b^2}-1 + j\)\,. 
\ee
In this limit, this relation implies $\beta_k = 0$ for all $k> 0$, with $\beta_0 = A_0$ the only non-vanishing term. As a consequence, the series sums into
\begin{align} \label{eq:resum}
G(1-w,1-\bar{w}) &\approx  i  C(\alpha_H,  \alpha_H, 2\alpha_L) \textrm{Res }  C(\alpha_L, \alpha_L, Q- 2\a_L) + \,  \mbox{continuous spectrum}\,.
\end{align}
Had we  worked at leading order throughout (i.e. at $O(c)$), and ignored the  $O(1)$  shifts in the infinite tower of poles (see \eqref{eq:polesRes} and \eqref{eq:hPm}), we would have obtained a divergent result due to the infinite sum of identical terms. However, the corrections arising from considering the $O(1)$ shifts in the weight of each term add up to make the leading order result finite. 
This  resummation is similar to the situation discussed in \cite{Heemskerk:2009pn} for four-point functions in holographic CFTs.  We will comment more on this point in the final discussion section.

The generalization  of this analysis to arbitrary insertion points is given in Appendix~\ref{app:doubletrace}. As can be read out from \eqref{eq:sumaribitrary},  it leads to 
\begin{align} 
G(1-w,1-\bar{w})  \approx  i & C(\alpha_H,  \alpha_H, 2\alpha_L) \textrm{Res }  C(\alpha_L, \alpha_L, Q- 2\a_L) \left| w \right|^{-2  \frac{\eta_L}{b^2}}    \( \frac{ |w|^{- \frac{\tilde \alpha}{2}} + |w|^{\frac{\tilde \alpha}{2}} }{2} \)^{- 4 \frac{\eta_L}{b^2}} \nonumber \\
&  +  \mbox{continuous spectrum} \, ,
\end{align}
or  in  terms of the original insertion variable $x$
 \begin{align} \label{eq:resumgen}
G(x,\bar{x})  \approx  i & C(\alpha_H,  \alpha_H, 2\alpha_L) \textrm{Res }  C(\alpha_L, \alpha_L, Q- 2\a_L) \left| 1-x\right|^{-2  \frac{\eta_L}{b^2}}    \( \frac{| 1-x|^{- \frac{\tilde \alpha}{2}} + |1-x|^{\frac{\tilde \alpha}{2}} }{2} \)^{- 4 \frac{\eta_L}{b^2}} \nonumber \\
&  +  \mbox{continuous spectrum} \, .
\end{align}
%

%%%%%%%%%%%%%%%%%%%%%%%%%%%%%%%%%%%%%%%%%%%%%%%
 \subsection{Semiclassical HHLL correlator }
\label{sec:SemiclassicalDOZZ}
To complete the computation of the discrete contributions to the correlator we need to evaluate the semiclassical limit of the factor 
\begin{align}
iC(\alpha_H,  \alpha_H, 2\alpha_L) &\textrm{Res }  C(\alpha_L, \alpha_L, Q- 2\a_L) =  \\
&\left[ \pi \mu \gamma(b^2) b^{2-2b^2}\right]^{1 + \frac{1-2\eta_H-2\eta_L}{b^2}}   \frac{\Upsilon_0 \Upsilon_b^2(2\alpha_H) \Upsilon_b(4\alpha_L)}{ \Upsilon_b (2\alpha_H + 2\alpha_L -Q) \Upsilon_b (2\alpha_H - 2\alpha_L) \Upsilon_b^2 (2\alpha_L )} \nonumber  \, . 
\end{align}
First  we use the recursion relations  \eqref{eq:upsilonminusb} and 
\be
\Upsilon_b\(x - \frac 1 b\) = \gamma \(\frac  x b - \frac{1}{b^2}\)^{-1} b^{1+\frac{2}{b^2} -\frac{2x}{b}} \Upsilon_b(x) 
\ee
to write
\begin{align}
iC&(\alpha_H,  \alpha_H, 2\alpha_L) \textrm{Res }  C(\alpha_L, \alpha_L, Q- 2\a_L) = \left[ \pi \mu \gamma(b^2)\right]^{1 + \frac{1-2\eta_H-2\eta_L}{b^2}} b^2   \\
& \times \gamma\(\frac{2\eta_H + 2\eta_L -1}{b^2}\) \gamma\(2\eta_H +2\eta_L-1 -b^2\)  \frac{\Upsilon_0 \Upsilon_b^2(2\alpha_H) \Upsilon_b(4\alpha_L)}{ \Upsilon_b (2\alpha_H + 2\alpha_L) \Upsilon_b (2\alpha_H - 2\alpha_L) \Upsilon_b^2 (2\alpha_L )} \,. \nonumber
\end{align}
Then to evaluate the limit of the $\Upsilon$ functions as $b \to 0$ we use the asymptotic formulae  \cite{Harlow:2011ny}
\bea
\Upsilon_0 &=& \frac{\upsilon}{\sqrt b}  e^{-\frac{1}{b^2} \[\frac{1}{4} \log b - F(0) +O(b^4 \log b)\]}\\
\Upsilon_b\(\frac \eta b\)& =& e^{\frac{1}{b^2} \[F(\eta) - (\eta -1/2)^2 \log b +O(b \log b)\]}
\eea
for $0< {\rm Re}(\eta) <1$, where
\be
F(\eta) \equiv \int_{1/2}^\eta \log \gamma(x) dx \,,
\ee
and $\upsilon$ is an $O(1)$ constant. 
To evaluate the $\gamma$ functions, we use the asymptotic expression 
\be
\G(x) = \left\{\begin{array}{cc} 
e^{x \log x - x + O(\log x)} & {\rm Re}\,(x) >0 \\
(e^{i \pi x} - e^{-i \pi x})^{-1} e^{x \log (-x) - x + O(\log(- x))} &  {\rm Re}\,(x) < 0
\end{array}\right. \quad \text{for~} |x|\to \infty\, , 
\ee
as well as the $\G$-function reflection formula.
At leading order in the semiclassical limit we obtain
\begin{align} \label{eq:DOZZsemiclassical}
&i  C(\alpha_H,  \alpha_H, 2\alpha_L) \textrm{Res }  C(\alpha_L, \alpha_L, Q- 2\a_L) \sim  \nonumber \\
&  \exp\[ -\frac{1}{b^2} \Big(   \(2 \eta_H+2 \eta_L-1\) \log  \lambda -   2 F(2\eta_H)  -  F(4\eta_L)   + F(2\eta_H + 2 \eta_L) + F(2 \eta_H - 2\eta_L)  \right.  \nonumber \\
&\left. \ \ \ \ \ \ \  + 2 F( 2\eta_L) - F(0) + 2  (1-2 \eta _H -2 \eta_L) \[  \log (1-2 \eta _H -2 \eta_L) -1\] \Big)  \]  \nonumber\\
&\times \frac{1}{e^{i \pi( 2 \eta_H + 2 \eta_L -1)/b^2} - e^{-i \pi( 2 \eta_H + 2 \eta_L -1)/b^2}}\,. 
\end{align}
At  the order we are working, this is actually completely determined by the semiclassical limit of the DOZZ coefficient $C(\alpha_H,  \alpha_H, 2\alpha_L)$, as Res $ C(\alpha_L, \alpha_L, Q- 2\a_L)$ only  contributes an $O(1)$ multiplicative factor which we are ignoring.

However we still have to interpret the last factor in \eqref{eq:DOZZsemiclassical}.  An important point is that when ${\rm Im} \,(2\eta_H +2\eta_L-1) =0$, as is the case here, this factor has oscillatory behavior and its semiclassical limit is ill-defined. As discussed in  \cite{Harlow:2011ny}, for non-vanishing imaginary part we could instead write
\be \label{eq:series}
\frac{1}{e^{i \pi( 2 \eta_H + 2 \eta_L -1)/b^2} - e^{-i \pi( 2 \eta_H + 2 \eta_L -1)/b^2}} = \pm \sum_{N=0}^\infty e^{\mp 2 \pi i (N+1/2)( 2 \eta_H + 2 \eta_L -1)/b^2}
\ee 
for $|e^{i \pi( 2 \eta_H + 2 \eta_L -1)/b^2}|  > 1$ or $<1$, respectively. This is the statement of  \cite{Harlow:2011ny} that the analytic continuation of the DOZZ formula to the regime of weights under consideration is given, depending on the sign of the imaginary part of $(2\eta_H +2\eta_L-1)$,  by two different infinite sums over complex saddlepoint solutions of the form \eqref{eq:ThreePoint} with three operator insertions.  The condition ${\rm Im} \,(2\eta_H +2\eta_L-1) =0$ defines a Stokes wall, along which the contributing saddles change discontinuously. 
In order to arrive at  a well-defined semiclassical result, when continuing in  $\eta_L$ we therefore need to give a prescription for how we approach the Stokes wall. 

With this 
in mind, in the following we allow either sign possibility from \eqref{eq:series} in our expression and linearize in $\eta_L$ all terms in the exponential,
\begin{align} 
  i  C&(\alpha_H,  \alpha_H, 2\alpha_L) \textrm{Res }  C(\alpha_L, \alpha_L, Q- 2\a_L)  \sim \nonumber \\
&   \sum_{N=0}^\infty  \exp\[ -\frac{1}{b^2} \Bigg(  \mp 2 \pi i \(N+\frac 1 2\)( 2 \eta_H + 2 \eta_L -1)  +  \(2 \eta_H+2 \eta_L-1\) \log  \lambda  + 4\eta_L \log 2  \right.  \nonumber \\
&\left. + 2  \[ (1-2 \eta _H -2 \eta_L) \log (1-2 \eta _H ) -(1-2 \eta _H)\]   \Bigg) \]\, .
\end{align}

The above result together with \eqref{eq:resumgen} gives the final expression for the contribution of the discrete terms entering in the block decomposition of the four-point function. We therefore have for the HHLL correlator 
\begin{align}
\langle  V_{\a_1}(1)   V_{\a_2}(\infty )  V_{\a_3}(0) V_{\a_4}(x,\bx)  \rangle &\approx \sum_{N=0}^\infty  e^{- \frac{\tilde S_N(x, \bar x) }{b^2}} + \,\mbox{continuous spectrum}\,, \label{eq:HHLLfinal}
\end{align}
where 
\begin{align} 
\tilde{S}{_N}(x, \bar x)  =& - (1- 2\eta_H- 2\eta_L) \log \lambda \pm 2\pi i \(N+\frac{1}{2}\) ( 1-2\eta_H - 2\eta_L) - 2 (1-2\eta_H) \nonumber \\
& + 2 (1-2\eta_H - 2\eta_L) \log(1-2\eta_H)  +4\eta_H(1-\eta_H)\log |z_\infty|  \nonumber \\ 
&+  4 \eta_L \log \( 1+ |1-x|^{1-2\eta_H}\) + 4 \eta_L \eta_H \log |1-x| + \dots ,
\end{align} 
which exactly matches the form of the saddles we have determined in \eqref{eq:superpmono}.   We keep either the plus sign or the minus sign.  This determines which half of the saddles in \eqref{eq:superpmono} should be summed over.

The analysis of the conformal block decomposition has therefore allowed us to determine which of the saddles we computed with different methods in the previous sections contribute to the correlator. Notice that here the multiple contributions labeled by $N$ arise from analytic properties  of the DOZZ coefficients when continued away from the normalizable regime, while in the previous sections they arose from the symmetry properties  of the Liouville equation.   
Comparison with the path integral results of those previous sections suggests that the sum over lighter discrete exchanges is the dominant contribution in the linearized HHLL correlator \eqref{eq:HHLLfinal},  and that the continuous spectrum only contributes at higher orders. However we have not explicitly evaluated the integral over the continuum of normalizable states. We thus cannot completely rule out the existence of other contributions to the linearized HHLL correlator, which, if present,  in the path integral approach could correspond to additional complex saddles with a different structure than those derived in Sec.~\ref{sec:linearized} and \ref{sec:mono}. In particular,  as the integral over the continuum of normalizable states can produce short distance singularities at $x \sim 0, 1, \infty$, as reviewed in \cite{Harlow:2011ny}, we cannot exclude the presence of singularities that were not captured by the saddle points we considered in our linearized path integral approach. In general we also expect short-distance singularities  to appear at higher orders in the conformal weight of the perturbatively heavy operators.

%%%%%%%%%%%%%%%%%%%%%%%%%%%%%%%%%%%%%%%%%%%%%%%%%%%%%%%%%%%%%%%%%%%%%%%%
%%%%%%%%%%%%%%%%%%%%%%%%%%%%%%%%%%%%%%%%%%%%%%%%%%%%%%%%%%%%%%%%%%%%%%%%
\section{Discussion} \label{sec:discussion}

We conclude this paper with three observations.
First, in Sec.~\ref{sec:light}, we will suggest a way of treating  perturbatively heavy operators as probes, in the same sense as conventional Liouville light operators.  This suggests an alternative way of approaching the computations in this paper.
Second, in Sec.~\ref{sec:lorentz}, we will draw a connection between our results and the structure of correlators in theories that have holographic duals  \cite{Heemskerk:2009pn}.  The connection is surprising given that Liouville theory is not known to enjoy a conventional holographic duality.
Third, in Sec.~\ref{sec:ententropy},   we will remark on the possibility of interpreting HHLL correlators as computing the entanglement entropy in excited states of Liouville theory.

%%%%%%%%%%%%%%%%%%%%%%%%%%%%%%%%%%%%%%%%%%%%%%%%%%%%%%%%%%%%%%%%%%%%%%%%
\subsection{Light operator insertions in a semiclassical background} \label{sec:light}

As reviewed around \eqref{eq:Lsemi} correlators involving heavy and ``Liouville light'' operators are computed by treating the latter in a probe approximation \cite{Zamolodchikov:1995aa}
\be
\langle  V_\frac{\eta_1}{b}(z_1,\bar{z}_1) \dots V_\frac{\eta_j}{b}(z_j,\bar{z}_j) V_{ \alpha_1}(w_1,\bar{w}_1)\dots V_{\alpha_n}(w_n,\bar{w}_n) \rangle \approx
  e^{-\frac{c}{6}\tilde{S}[\phi_c] } \prod_{i=1}^{n} e^{ \frac{\alpha_{i}}{b} \phi_c(w_i,\bar{w}_i) }  \, , \label{eq:lightonH}
\ee
where the  Liouville field $\phi_c$ solves the Liouville equation in presence of the heavy insertions with momenta $\eta_k/b$, and  $\alpha_i$ are the  momenta of the light operators. The right hand side of \eqref{eq:lightonH}  is understood to imply an integral over the moduli  of $\phi_c$, which leave the action invariant, as well as a sum over different contributing saddles.

Let us assume that the probe approximation extends to our correlator, which involves perturbatively heavy operators with $\alpha_L = \eta_L/b$ and $\eta_L\ll1$ fixed as $b\rightarrow 0$.  The complex HH background Liouville solutions $\phi_c$ are related to each other by $2\pi i N$ shifts and are given in \eqref{eq:phiHW}. Each of these complex solutions has a modulus $\kappa$ taking values in the upper half plane with the real axis removed, which we here denote as  $\mathbbm{H}$.  In the probe approximation, we therefore evaluate
\begin{align}
\langle  V_\frac{\eta_H}{b}(z_1,\bar{z}_1) V_\frac{\eta_H }{b}(z_2,\bar{z}_2) &V_\frac{\eta_L }{b}(z_3,\bar{z}_3)  V_\frac{\eta_L }{b}(z_4, \bar{z}_4) \rangle   \approx \label{eq:ActionLights}\\ 
&\approx  \sum_{N \in \mathcal{T}}  e^{-\frac{1}{b^2}\( \tilde{S}_N^{(0)} -  4   \pi i N \eta_L  \)}   \int_{\mathbbm{H}}  e^{\frac{\eta_L}{b^2}\phi_{c}(z_3,\bar{z}_3)}e^{\frac{\eta_L}{b^2}\phi_{c}(z_4,\bar{z}_4)}d\kappa\wedge d\bar{\kappa}\,   \nonumber \\
&\approx   \sum_{N \in \mathcal{T}}  e^{-\frac{1}{b^2}\( \tilde{S}_N^{(0)} -  4   \pi i N \eta_L   + 2 \eta_L \log \l\)}    \int_{\mathbbm{H}} \frac{ \kappa^{\frac{4 \eta_L}{b^2}} d\kappa \wedge d\bar{\kappa}}{ \[ \( \zeta_3 \kappa^2- \chi_3\) \( \zeta_4 \kappa^2- \chi_4 \)\]^{\frac{2\eta_L}{b^2}} } \,  ,  \nonumber
\end{align}
where $\mathcal{T}$ denotes the set of contributing saddles, $\tilde{S}_N^{(0)} $ is the HH on-shell action given in \eqref{eq:St0} and  we have  introduced the notation
\be
\zeta_i \equiv \left| z_{1i}\right|^{2\eta_H}\left| z_{2i}\right|^{2-2\eta_H}\, , \qquad \qquad \chi_i  \equiv \frac{\left| z_{1i}\right|^{2-2\eta_H}\left| z_{2i}\right|^{2\eta_H}}{ \left| z_{12}\right|^2 \( 1-2\eta_H\)^2}  \, .
\ee
Using Stokes' theorem,
the integral over the upper half plane can be reduced to  a line integral  over $\mathbbm{R} + i \epsilon $
\begin{align} \label{eq:kint}
\langle  V_\frac{\eta_H}{b}&(z_1,\bar{z}_1)  V_\frac{\eta_H }{b}(z_2,\bar{z}_2)  V_\frac{\eta_L }{b}(z_3,\bar{z}_3)  V_\frac{\eta_L }{b}(z_4, \bar{z}_4) \rangle   \approx   \\
 &\approx  \sum_{N \in \mathcal{T}}  e^{-\frac{1}{b^2}\( \tilde{S}_N^{(0)} -  4   \pi i N \eta_L   + 2 \eta_L \log \l\)}     \int_{-\infty }^{ \infty} \frac{ \( q-i\epsilon\)  \(q+i\epsilon\)^{\frac{4 \eta_L}{b^2}} d q}{\[ \( \zeta_3 \(q+i\epsilon\)^2- \chi_3\) \( \zeta_4 \(q+i\epsilon\)^2- \chi_4 \)\]^{\frac{2\eta_L}{b^2}} }  \, .\nonumber
\end{align}
We can give an estimate of  this integral in the limit $b \to 0$ by considering its saddlepoints. At leading order in $b\to 0 $, the saddles are the four roots of 
\be
(q+ i \eps)^4 =  \frac{ \chi_3 \chi_4 }{\zeta_3 \zeta_4} \,  . \label{eq:saddles}
\ee
While it would be interesting to study systematically which saddlepoints contribute, here we simply  notice that if we assume  that only  the two saddles 
\be
q + i \eps = \pm i \[ \frac{  \chi_3 \chi_4} {\zeta_3 \zeta_4}\]^{\frac{1}{4}}\,  \label{eq:saddles}
\ee
(for which the integral evaluates to the same function) can contribute, we then find
\begin{align} 
\langle  V_\frac{\eta_H}{b}(1) &V_\frac{\eta_H }{b}(\infty) V_\frac{\eta_L }{b}(0)  V_\frac{\eta_L }{b}(x, \bx) \rangle       \\ &\approx\sum_{N \in \mathcal{T}}  e^{-\frac{1}{b^2}\tilde{S}_N^{(0)}  }   e^{ -\frac{2 \eta_L}{b^2} \( - 2  \pi i  (N  + \frac{1}{2}) + \log \lambda - 2 \log (  1-2 {\eta_H} )  + 2  \log  (  1+  |1-x|^{1- 2 \eta_H})   +2 \eta_H\log |1-x|     \)  } \nonumber
\end{align}
for canonical insertion points $z_1 = 1$, $z_2 = \infty$, $z_3 = 0$, $z_4 = x $. 

This is the same result we have found before. 
At least in principle, through a careful steepest descent analysis on the full moduli space $\mathbbm{H} \times \mathbbm{Z}$, it should be possible to also determine the  set of contributing  saddles $\mathcal{T} $  and to verify that this is in agreement with the result of Sec.~\ref{sec:DOZZ}.

%%%%%%%%%%%%%%%%%%%%%%%%%%%%%%%%%%%%%%%%%%%%%%%%%%%%%%%%%%%%%%%%%%%%%%%%%
\subsection{Lorentzian singularity}
 \label{sec:lorentz} 
  
In \cite{Heemskerk:2009pn} it was conjectured that any CFT with a large-$N$ expansion and a large gap in the spectrum of low-dimension operators has a local bulk dual. 
The original analysis of \cite{Heemskerk:2009pn} considers a low-dimensional spectrum of operators that, to leading non-trivial order in a $1/N$ expansion, have a closed algebra among themselves.  These are the identity, a unique single-trace operator $\mathcal{ O}$ of conformal dimension $\Delta$ (assumed to be $O(1)$ in the large $N$ limit) and double-trace operators
\be \label{eq:doubletr}
\mathcal{ O}_{m,l} \equiv \mathcal{ O} \, \overset{\leftrightarrow}{\partial_{\mu_1}}\dots\overset{\leftrightarrow}{\partial_{\mu_\ell}} \(\overset{\leftrightarrow}{\partial_{\nu}} \,\overset{\leftrightarrow}{\partial^{\nu}} \)^m  \mathcal{ O} - \, \textrm{traces}\,,
\ee
with spin $\ell$ and conformal dimension $\Delta_{m,\ell} = 2\Delta + 2m + \ell + O(1/N^2)$. Imposing  $\mathbb{Z}_2$ symmetry also  guarantees  that $\mathcal{ O}$ does not itself appear in the $\mathcal{ O}\mathcal{ O}$ OPE. 
Under these assumptions, in a $1/N$ expansion the four-point function of operators $\mathcal{O}$ contains a singularity  after continuation to Lorentzian signature that is not present in single conformal blocks, but arises from  resummation of double-trace exchanges \cite{Heemskerk:2009pn}. 
This singularity was associated to a notion of bulk locality, since it occurs when the operators on the boundary are aligned to give rise to a local scattering process in a dual AdS bulk \cite{Gary:2009ae,Heemskerk:2009pn}.

Even though there are important differences in the structure of the correlator we are studying,\footnote{In particular, we are considering a four-point function of two pairs of operators of conformal dimension  $O(c)$ instead of a four-point function of identical operators of conformal dimension $O(1)$. Importantly the identity exchange does not contribute in our expansion,  and the lowest states exchanged are those with conformal weights \eqref{eq:hPm}.} the conformal block analysis we presented in Sec.~\ref{sec:DOZZ}  has similarities to the one of \cite{Heemskerk:2009pn}. 
In fact,  we can see that the conformal weights $h_{P_m}$ of the discrete contributions appearing in the decomposition of the Liouville HHLL correlator, given in formula \eqref{eq:hPm}, coincide with those of spinless ``double-trace'' operators in \eqref{eq:doubletr}.
Specifically, taking $ V_{\alpha_{P_0}} =  e^{\frac{\eta_L}{b^2} \phi_c}$ and using the Liouville equation \eqref{eq:Lsphere0}, the combination
\be
V_{\alpha_{P_0}} \( \overset{\leftrightarrow}\del_z  \overset{\leftrightarrow}\del_{\bar z} \)^m V_{\alpha_{P_0}} \, 
\ee 
 can be identified  to leading order in  $b\to 0$ with $V_{\a_{P_{m}}}$   of momentum $\alpha_{P_m} = 2 \alpha_L+ mb$.

Intriguingly, we find that these contributions resum to a leading semiclassical result that contains a peculiar Lorentzian singularity, as in \cite{Heemskerk:2009pn}.
Specifically, transforming the result in Sec.~\ref{sec:SemiclassicalDOZZ}  to the cylinder, $x=1-e^{\tau + i\theta}$, and Wick rotating to Lorentzian time we have
\begin{equation}
\langle V_{\frac{\eta_H}{b}} (-\infty) V_{\frac{\eta_L}{b}}(0,0) V_{\frac{\eta_L}{b}}(t,\theta) V_{\frac{\eta_H}{b}}(\infty )\rangle_{\text{cyl}} \sim \cos \( \frac{\tilde{\alpha} t}{2}\)^{-\frac{4\eta_L}{b^2}} \,. \label{eq:lorentz}
\end{equation}
The correlator exhibits a singularity at $t=\pi/\tilde{\alpha}$. In a hypothetical bulk dual, one would expect the heavy operators to create a conical defect spacetime with opening angle $2\pi\tilde\alpha$, and $t=\pi/\tilde{\alpha}$ is precisely the time it would take a massless excitation to reach the defect and return to the boundary.   

The result summarized in \eqref{eq:resumgen}, which eventually led to \eqref{eq:lorentz}, was extracted  by  evaluating the integral representation of a specific hypergeometric function in the saddle point approximation (see Appendix \ref{app:doubletrace}). In particular,  out of two possible saddles, only one contributed in  the Euclidean regime. However, when continuing to Lorentzian time and following the evolution of the HHLL correlator, one can check  that a transition between saddles takes place at  $t=\pi/\tilde{\alpha}$, similarly to what was observed in \cite{Gary:2009ae,Heemskerk:2009pn}. Interestingly, the functional dependence of the HHLL correlator past the singularity can be obtained from the analysis of Sec.~\ref{sec:mono} in terms of the accessory parameter that would correspond to solutions with $SU(1,1)$ monodromy. At the level of the  Euclidean  analysis, such a choice was not justified by any reality or single-valuedness requirement.  It would be interesting to understand how the continuation to Lorentzian signature and the associated appearance of singularities that are absent in the Euclidean HHLL correlator are reflected in and modify the monodromy method of Sec.~\ref{sec:DOZZ}.

%%%%%%%%%%%%%%%%%%%%%%%%%%%%%%%%%%%%%%%%%%%%%%%%%%%%%%%%%%%%%%%%%%%%%%%%
\subsection{Entanglement entropy in heavy excited states}
\label{sec:ententropy}

In conformal field theory at large $c$, results for Euclidean correlation functions  involving perturbatively heavy operators have been used to evaluate correlators of R\'enyi replica twists, which compute entanglement entropy and other entanglement related measures (see e.g. \cite{Hartman:2013mia,Asplund:2014coa,Kulaxizi:2014nma}).

In particular, HHLL correlators  in holographic CFTs have been directly related  to single interval entanglement entropy in excited eigenstates and have been evaluated under the assumption that the correlator, and thus the entanglement entropy, are dominated by the semiclassical Virasoro vacuum block \cite{Asplund:2014coa}.  The insertions of heavy primaries are  interpreted as preparing the system in an excited eigenstate of the Hamiltonian and light (perturbatively heavy) insertions are identified with replica twists in the limit in which they compute the entanglement entropy.\footnote{R\'enyi replica twists transform like primaries of conformal weight $h_n=\bar h_n = \frac{c}{24}\(n-\frac{1}{n}\)$, and in the replica approach the entanglement entropy is obtained evaluating correlators involving twist fields for each integer $n \ge 2$  and analytically continuing the result to $n=1$. In this sense, twist operators in the limit in which they compute the entanglement entropy have weight that scales with $c$ but such that $h_n/c \ll1$, and can be treated as being perturbatively heavy.}

In Liouville theory, there are subtle issues regarding the normalizability of states, and,  as we showed, there is no identity contribution to the HHLL 4-point.   Thus it is not completely clear whether HHLL correlators can be repurposed to compute entanglement entropies.
Intriguingly, if such an interpretation were possible, our path integral analysis would predict that the single interval entanglement entropy for a Liouville excited state does not depend on the size of the interval.  This also happens in quantum field theories with a mass gap, for interval sizes larger than the correlation length (see e.g. \cite{Cardy:2007mb}).

%%%%%%%%%%%%%%%%%%%%%%%%%%%%%%%%%%%%%%%%%%%%%%%%%%%%%%%%%%%%%%%%%%%%%%%%

\section*{Acknowledgments}

We thank Lorenzo Di Pietro, Laura Donnay, Davide Gaiotto, Jaume Gomis, Daniel Harlow, Erik Tonni and Alexander Zamolodchikov for useful discussions.
This work was supported in part by a grant from the Simons Foundation (\#385592, Vijay Balasubramanian) through the It From Qubit Simons Collaboration, by the Belgian Federal Science Policy Office through the Interuniversity Attraction Pole P7/37, by FWO-Vlaanderen through projects G020714N, G044016N and Odysseus grant G.001.12, by the European Research Council grant no. ERC-2013-CoG 616732 HoloQosmos, by COST Action MP1210 The String Theory Universe, and by Vrije Universiteit Brussel through the Strategic Research Program ``High-Energy Physics''. Research at Perimeter Institute is supported by the Government of Canada through Industry Canada and by the Province of Ontario through the Ministry of Research \& Innovation. A.B. and F.G. thank the Galileo Galilei Institute for Theoretical Physics (GGI) and SISSA for the hospitality during completion of this work, within the program ``New Developments in AdS$_3$/CFT$_2$ Holography'', as well as INFN and ACRI (Associazione di Fondazioni e di Casse di Risparmio S.p.a.) for partial support. T.D.J.\ is Aspirant FWO-Vlaanderen.

%%%%%%%%%%%%%%%%%%%%%%%%%%%%%%%%%%%%%%%%%%%%%%%%%%%%%%%%%%%%%%%%%%%%%%%%%

\appendix
 
\section{Computation of the monodromy matrices}  \label{app:monocomp}
 
In this section we collect the details of the computation of the monodromy of the solutions \eqref{psi1}-\eqref{psi2}
\bea
\psi_1(z) &=& (1-z)^{\frac{1+ \tilde\alpha}{2}} + \eta_L \Bigg[ (1-z)^{\frac{1+\tilde\alpha}{2}} \frac{\left(\frac{\cx}{\eta_L} (1-x)+1 \right)\log \frac{z}{z-x} + \frac{(x-2)z+x}{z(z-x)}}{\tilde\alpha}  \label{app:psi1}\\ 
&& + (1-z)^{\frac{1- \tilde\alpha}{2}} \int dz~ \frac{(1-z)^{\tilde\alpha} \left( \frac{\cx(x-1)x z (x-z)}{\eta_L} - x^2 (z+1)+ 2 x z (z+1)-2z^2 \right)}{z^2 \tilde\alpha(x-z)^2}\Bigg] \nonumber \,  \\
\psi_2(z) &=& (1-z)^{\frac{1- \tilde\alpha}{2}} - \eta_L \Bigg[  (1-z)^{\frac{1- \tilde\alpha}{2}} \frac{\left(\frac{\cx}{\eta_L} (1-x)+1 \right)\log \frac{z}{z-x} + \frac{(x-2)z+x}{z(z-x)}}{\tilde\alpha} \label{app:psi2} \\
&&  +(1-z)^{\frac{1+\tilde\alpha}{2}} \int dz~ \frac{(1-z)^{-\tilde\alpha} \left( \frac{\cx(x-1)x z (x-z)}{\eta_L} - x^2 (z+1)+ 2 x z (z+1)-2z^2 \right)}{z^2 \tilde\alpha(x-z)^2}  \Bigg]  \nonumber \, . 
\eea
To evaluate the monodromy around  $z=0$, as in Appendix D of \cite{Fitzpatrick:2014vua}, we just notice that  in the first line of \eqref{app:psi1} and of  \eqref{app:psi2}  the only non-trivial contribution comes from $\log z  \to \log z  +  2 \pi i $.  The monodromy of the expression in  the second line of \eqref{app:psi1} and of  \eqref{app:psi2} is evaluated through the residues of the integral. For the monodromy around $z=0$ we have
\begin{align}
\psi_1 &\to  \psi_1   +  \frac{2 \pi i}{\tilde \a }  \eta_L \left[ (1-z)^{\frac{1+ \tilde\alpha}{2}}  \(\frac{\cx}{\eta_L}(1-x)+1\) +   (1-z)^{\frac{1- \tilde\alpha}{2}}  \(-\frac{\cx}{\eta_L}(1-x)- 1+\tilde\a  \) \right]\\
\psi_2 & \to  \psi_2   - \frac{2 \pi i}{\tilde \a }  \eta_L \left[ (1-z)^{\frac{1- \tilde\alpha}{2}}  \(\frac{\cx}{\eta_L}(1-x)+1\) +   (1-z)^{\frac{1+ \tilde\alpha}{2}}  \(-\frac{\cx}{\eta_L}(1-x)- 1-\tilde\a \) \right]
\end{align}
At linear order in $\eta_L$, the monodromy matrix is
\be
 M_{\gamma_0} = \Id+ \frac{2\pi i}{\tilde\alpha}
 \left(\begin{array}{cc} \cx(1-x) + \eta_L & -\cx(1-x) - \eta_L (1-\tilde\alpha) \\
\cx(1-x) + \eta_L (1+\tilde\alpha)&  -\cx(1-x) - \eta_L\end{array}\right)  \, .
\ee
The monodromy around $z=x$ is computed analogously and the linearized monodromy matrix is given by
\be
 M_{\gamma_x} = \Id + \frac{2\pi i}{\tilde\alpha}
 \left(\begin{array}{cc} -\cx(1-x) - \eta_L & (1-x)^{\tilde\alpha} \(  \cx(1-x) + \eta_L (1+\tilde\alpha) \) \\
 (1-x)^{-\tilde\alpha} \( -\cx(1-x) - \eta_L (1-\tilde\alpha) \)   &  \cx(1-x) +\eta_L \end{array}\right) \, .
\ee
To compute the monodromy  around $z=1$ in the complex plane, notice that the expressions in the first line of \eqref{app:psi1} and \eqref{app:psi2} simply pick up a phase $e^{ (1+ \tilde\alpha) \pi i }$  and $e^{ (1- \tilde\alpha) \pi i}$  respectively. To determine the monodromy of the expression in the second line  \eqref{app:psi1} and \eqref{app:psi2},  it is more  convenient to explicitly perform  the integrals and express them in terms of hypergeometric functions. The contribution entering in $\psi_1$ can be written as 
\begin{align}
&\int dz \frac{(1-z)^{\tilde\alpha} \left( \frac{\cx(x-1)x z (x-z)}{\eta_L} - x^2 (z+1)+ 2 x z (z+1)-2z^2 \right)}{z^2 \tilde\alpha (x-z)^2} =  \nonumber\\
&\frac{(1-z)^{1+\tilde\alpha}}{\tilde\alpha z}  + \frac{(1-z)^{1+\tilde\alpha}  \hf[2,1+\tilde\alpha, 2+\tilde\alpha, \frac{1-z}{1-x}]}{\tilde\alpha(1+\tilde\alpha) (1-x)}  +\frac{(1-z)^{1+\tilde\alpha} (\frac{\cx}{\eta_L} (1-x)+1) \hf[1,1,1-\tilde\alpha, \frac{1-x}{z-x}] }{\tilde\alpha^2 (x-z)}  \nonumber \\
&- \frac{(-z)^{\tilde\alpha} (\frac{\cx}{\eta_L} (1-x)+1 -\tilde\alpha)\hf[-\tilde\alpha, -\tilde\alpha, 1-\tilde\alpha, \frac 1 z] }{\tilde\alpha^2}    \, .  \label{eq:intpsi1}
\end{align}%
The monodromy of the the first two terms  is completely determined by the factor $(1-z)^{1+\tilde\alpha}$, as there is no branch cut crossing in the hypergeometric function. 
For the other two terms, the monodromy is determined by the properties of the hypergeometric functions. In order to extract their monodromy we just notice that 
\bea
\hf\[1,1,1-\tilde\alpha, \frac{1-x}{z-x}\] & = & \hf\[1,1,1-\tilde\alpha, 1  - \frac{z-1}{z-x}\]  \, ,  \\
\hf\[1,1,1-\tilde\alpha, \frac{1}{z}\] & = & \hf\[1,1,1-\tilde\alpha, 1  - \frac{z-1}{z}\]   \, ,
\eea
and use the identity 
\begin{align}
\hf[a,b,c,1-q]  = \frac{\Gamma[c]\Gamma[c-a-b]}{\Gamma[c-a]\Gamma[c-b]} \hf[a,b,a+b-c+1, q]& \\
 + q^{c-a-b} \frac{\Gamma[c]\Gamma[a+b-c]}{\Gamma[a]\Gamma[b]}& \hf[c-a,c-b,c-a-b+1,q] \nonumber
\end{align}
to isolate, in the prefactors of the form $q^{c-a-b}$, all the contributions with a non-trivial monodromy as  $z$ is taken in a loop around $z=1$ (which corresponds to taking  $q$ around 0). 
This gives
\bea
\hf\[1,1,1-\tilde\alpha, \frac{1-x}{z-x}\] & = & \frac{\tilde\alpha}{1+\tilde\alpha} \hf\[1,1,2+\tilde\alpha, \frac{1-z}{x-z}\] +
\frac{\pi \tilde\alpha}{\sin \pi \tilde\alpha} \frac{z-x}{1-x} \left(\frac{1-x}{z-1}\right)^{1+\tilde\alpha}  \\
\hf\[1,1,1-\tilde\alpha, \frac{1}{z}\] & = & \frac{\tilde\alpha}{1+\tilde\alpha} \hf\[1,1,2+\tilde\alpha,\frac{z-1}{z}\] +
\frac{\pi \tilde\alpha}{\sin \pi \tilde\alpha} z \left(\frac{1}{z-1}\right)^{1+\tilde\alpha}  
\eea
so that the integral  in \eqref{eq:intpsi1} reduces to
\begin{align}
& \frac{(1-z)^{1+\tilde\alpha}}{\tilde\alpha} \Bigg\{ \frac{1}{z} + \frac{ \hf[2,1+\tilde\alpha, 2+\tilde\alpha, \frac{1-z}{1-x}]}{(1+\tilde\alpha) (1-x)}  +\frac{\frac{\cx}{\eta_L} (1-x)+1 }{(1+\tilde\alpha) (x-z)} \hf\[1,1,2+\tilde\alpha, \frac{1-z}{x-z}\]  \nonumber \\
 &+   \frac{ \frac{\cx}{\eta_L} (1-x)+1- \tilde\alpha}{ (1+\tilde\alpha) z} \hf\[1,1,2+\tilde\alpha, \frac{z-1}{z}\] \Bigg\}  \nonumber \\
& - \frac{\pi e^{ \tilde\alpha \pi i } }{\tilde\alpha \sin \pi \tilde\alpha}  \left[ \left(\frac{\cx}{\eta_L} (1-x)+1 \right) \(1-(1-x)^{\tilde\alpha}\) -\tilde\alpha \right] \,.
\end{align}
Using this expression  we can  schematically rewrite $\psi_1(z)$ by isolating those terms that have non-trivial mondromy
\be
\psi_1(z)  = (1-z)^{\frac{1+ \tilde\alpha}{2}} + \eta_L  (1-z)^{\frac{1+ \tilde\alpha}{2}}  {\rm mon}_{11} + \eps_{L}  (1-z)^{\frac{1- \tilde\alpha}{2}} {\rm mon_{12}}
\ee
where
\begin{align}
 & {\rm mon}_{11} = \frac{\left(\frac{\cx}{\eta_L} (1-x)+1 \right)\log \frac{z}{z-x} + \frac{(x-2)z+x}{z(z-x)}}{\tilde\alpha} + \frac{ (1-z)}{\tilde\alpha} \Bigg[ \frac{1}{z} + \frac{ \hf[2,1+\tilde\alpha, 2+\tilde\alpha, \frac{1-z}{1-x}]}{(1+\tilde\alpha) (1-x)}  \nonumber \\
&+ \frac{ \frac{\cx}{\eta_L} (1-x)+1 }{(1+\tilde\alpha) (x-z)} \hf\[1,1,2+\tilde\alpha, \frac{1-z}{x-z}\]  + \frac{ \frac{\cx}{\eta_L} (1-x)+1- \tilde\alpha}{(1+\tilde\alpha) z} \hf\[1,1,2+\tilde\alpha, \frac{z-1}{z}\] \Bigg]  \nonumber \\ 
&{\rm mon}_{12}  = - \frac{\pi e^{\tilde\alpha \pi i } }{\tilde\alpha \sin \pi \tilde\alpha} \left[ \left(\frac{\cx}{\eta_L} (1-x)+1 \right) \(1-(1-x)^{\tilde\alpha}\) -\tilde\alpha \right] \, .
\end{align}
By noticing that  $\psi_2(z)[ \tilde\alpha] =\psi_1(z)[ -\tilde\alpha]$ and  again writing schematically 
\be
\psi_2(z)  = (1-z)^{\frac{1- \tilde\alpha}{2}} + \eta_L (1-z)^{\frac{1- \tilde\alpha}{2}}~{\rm mon}_{22} + \eps_{L} (1-z)^{\frac{1+ \tilde\alpha}{2}}~{\rm mon}_{21}
\ee
with ${\rm mon}_{22}[\tilde \alpha] = {\rm mon}_{11} [-\tilde \a] $ and ${\rm mon}_{21}[\tilde \a] = {\rm mon}_{12}[-\tilde \a]$, it is now straightforward to write the monodromy transformation of $\{\psi_1,\psi_2 \}$ 
\begin{align}
\psi_1 &\to e^{\pi i (1+\a)}\psi_1  + \(  e^{\pi i (1-\a)}   - e^{\pi i (1+\a)}\) (1-z)^{\frac{1- \tilde\alpha}{2}}~{\rm mon}_{12} \\
\psi_2 & \to e^{\pi i (1-\a)}\psi_2  + \(  e^{\pi i (1+\a)}   - e^{\pi i (1-\a)}\) (1-z)^{\frac{1+ \tilde\alpha}{2}}~{\rm mon}_{21} \, , 
\end{align}
from which we immediately  read out the corresponding monodromy matrix at linear order in $\eta_L$
\begin{align}
& M_{\gamma_1} =- \left(\begin{array}{cc} e^{\tilde\alpha\pi i} & 0 \\0 & e^{-\tilde\alpha \pi i}\end{array}\right) + \\
&\frac{2\pi i}{\tilde\alpha}   
 \footnotesize{ 
 \left(\begin{array}{cc} 0 & e^{ \tilde\alpha \pi i}\left[  - \left(\cx (1-x)+\eta_L \right) \(1-(1-x)^{\tilde\alpha}\) + \eta_L \tilde\alpha\right] \\ 
e^{- \tilde\alpha \pi i}\left[ \left( \cx (1-x)+ \eta_L \right) \(1-(1-x)^{-\tilde\alpha} \)+ \eta_L \tilde\alpha\right]  & 0
\end{array}\right).  \nonumber
 }
\end{align}
%

%%%%%%%%%%%%%%%%%%%%%%%%%%%%%%%%%%%%%%%%%%%%%%%%%%%%%%%%%%%%%%%%%%%%%%%%%

\section{Sum over discrete exchanges for arbitrary insertions} \label{app:doubletrace}
 
In this section we generalize to arbitrary insertions the result of Sec.~\ref{sec:discrete}  for the sum over discrete terms entering the block decomposition of the four-point function.  

We start from an expression for $G(1-w, 1- \bw)$ analogous to \eqref{eq:discrunit}, but with $w$ representing an arbitrary insertion point instead of lying on the unit circle:
\begin{align} \label{eq:discrgeneric}
G &\approx  i \sum_{m=0}^\infty C(\alpha_H,  \alpha_H, 2\alpha_L + mb ) \textrm{Res }  C(\alpha_L, \alpha_L, Q- 2\a_L - mb) \left| w \right|^{2 (\tilde \alpha -1) h_L}
\left|1-w^{\tilde \alpha}\right|^{2m}  \tilde \a^{-2m}\nonumber\\
& \times F\left[ h_{P_m}, h_{P_m}, 2 h_{P_m}, 1-w^{\tilde\a} \right] F\left[ h_{P_m}, h_{P_m}, 2 h_{P_m}, 1-\bar w^{\tilde \a}  \right] + \mbox{continuous spectrum}\,  . 
\end{align}
Following \cite{Dolan:2000ut} and introducing the notation $u\equiv \left|1-w^{\tilde \alpha}\right|^2, v\equiv \left|w\right|^{2\tilde \alpha}$,  we can write the product of hypergeometrics 
as
\begin{equation}
F(\hP m , \hP m ,2 \hP m ,1-w^{\tilde\alpha})F(\hP m , \hP m ,2 \hP m ,1-\bar{w}^{\tilde\alpha}) = \sum\limits_{r,s=0}^{\infty}  \frac{(\hP m)^2_r ( \hP m)^2_{r+s}}{r!s! ( 2 \hP m)_r (2 \hP m)_{2r+s}} u^r (1-v)^s.
\end{equation}
Focusing on the sum over discrete terms (and temporarily  discarding the overall factor  $i \left| w \right|^{2 (\tilde \alpha -1) h_L}$)
\begin{align}
&\sum_{m,r,s=0}^\infty C(\alpha_H,  \alpha_H, 2\alpha_L + mb ) \textrm{Res }  C(\alpha_L, \alpha_L, Q- 2\a_L - mb)
    \frac{ u^{m+r}  (1-v)^s  \tilde \a^{-2m}(\hP m)^2_r ( \hP m)^2_{r+s}}{r!s! ( 2 \hP m)_r (2 \hP m)_{2r+s}}  \nonumber \\
&= \sum\limits_{s,k=0}^{\infty}u^k \tilde  \alpha^{-2k} \frac{(1-v)^s}{s!} \sum\limits_{r=0}^k A_{k-r}  \frac{\tilde \alpha^{2r}(\hP {k-r})_r^2 (\hP {k-r})_{r+s}^2}{r! (2\hP {k-r})_r (2\hP {k-r})_{2r+s}} \nonumber \\
&= \sum\limits_{s,k=0}^{\infty}   u^k \tilde  \alpha^{-2k} \frac{ (\hP k)_s^2 }{ (2 \hP k)_s }  \frac{(1-v)^s}{s!} \sum\limits_{r=0}^k  A_{k-r} \( \frac{\tilde \alpha}{2}\)^{2r} \frac{  (\hP {k-r})_r^3 }{r! (2\hP {k-r})_r \( \hP{k-r}+\frac{1}{2}\)_r}\\
&= \sum\limits_{k=0}^{\infty}    u^k\tilde  \alpha^{-2k}  \hf[\hP k, \hP k , 2 \hP k, 1-v ] \sum\limits_{r=0}^k  A_{k-r}\( \frac{\tilde \alpha}{2}\)^{2r}\frac{  (\hP {k-r})_r^3 }{r! (2\hP {k-r})_r \( \hP{k-r}+\frac{1}{2}\)_r} \nonumber 
\end{align}
where $ A_j \equiv  C (\alpha_H,  \alpha_H, 2\alpha_L + j b ) \textrm{~Res~}C(\alpha_L, \alpha_L, Q- 2\a_L - jb)$ as in \eqref{eq:AmDOZZ} in the main text.  In going from the second to the third line we used the definition of the Pochhammer symbol in terms of $\Gamma-$functions and their identities. The internal finite sum over  $r$ is the coefficient $\beta_k$ we defined in \eqref{eq:betak}. These are all vanishing except for $k=0$ and therefore,  remembering $\hP m \approx 2 h_L +  m \approx \frac{2 \eta_L}{b^2} +m $, 
\begin{align}
 \sum_{m,r,s=0}^\infty &C(\alpha_H,  \alpha_H, 2\alpha_L + mb ) \textrm{Res }  C(\alpha_L, \alpha_L, Q- 2\a_L - mb)
    \frac{ u^{m+r}  (1-v)^s  \tilde \a^{-2m}(\hP m)^2_r ( \hP m)^2_{r+s}}{r!s! ( 2 \hP m)_r (2 \hP m)_{2r+s}}  \nonumber \\
&= A_{0}  \,  \hf[ 2 h_L , 2h_L , 4  h_L, 1-v ]     \, . 
\end{align}
As a last step, we would like to take the semiclassical limit of this hypergeometric. For that we use the integral representation of the hypergeometric function
\begin{equation}
{}_2 F_1 [ 2 h_L,2h_L,4h_L,1-v] = \frac{\Gamma(4h_L)}{\Gamma(2h_L)^2} \int_1^{\infty} ds \, s^{-2h_L} \( s-1\)^{2h_L -1} \( s-1+v\)^{-2h_L}. \label{eq:intF}
\end{equation}
As $b\to 0$, the asymptotic expression for the gamma functions gives
\begin{align}
\frac{\Gamma (4h_L)}{\Gamma(2h_L)^2} \approx e^{4h_L\ln 2}\, .
\end{align}
The integral over $s$ can be evaluated in a saddlepoint approximation, but there is a branch cut  between $v=0$ and $v=-\infty$, which becomes relevant when analytically continuing  from  Euclidean to Lorentzian time.
  
There are in fact two real saddles at $s_{\pm}=1\pm \sqrt v$. 
In the Euclidean regime, $v >0$  and we take the $s_+$ saddle
\be
 \int\limits_1^{\infty} ds \, s^{-2h_L} \( s-1\)^{2h_L -1} \( s-1+v\)^{-2h_L} \approx e^{-2 h_L \log(1 + \sqrt v)^2  }
\ee
giving
\begin{align}
{}_2 F_1 [ 2h_L,2h_L,4h_L,1-v] \approx    e^{-4h_L \log \frac{ 1+ |w|^{\tilde \alpha} }{2}   }.
\end{align}
Using this result and   reinstating the factor $i \left| w \right|^{2 (\tilde \alpha -1) h_L}$ we had temporarily dropped, we obtain
\begin{align} 
&  i \sum_{m=0}^\infty C(\alpha_H,  \alpha_H, 2\alpha_L + mb ) \textrm{Res }  C(\alpha_L, \alpha_L, Q- 2\a_L - mb) \left| w \right|^{2 (\tilde \alpha -1) h_L}
\left|1-w^{\tilde \alpha}\right|^{2m}  \tilde \a^{-2m}  \nonumber\\
& \times F\left[ h_{P_m}, h_{P_m}, 2 h_{P_m}, 1-w^{\tilde\a} \right] F\left[ h_{P_m}, h_{P_m}, 2 h_{P_m}, 1-\bar w^{\tilde \a}  \right] \approx  i A_0 \left| w\right|^{-2  h_L}    \( \frac{ |w|^{- \frac{\tilde \alpha}{2}} + |w|^{\frac{\tilde \alpha}{2}} }{2} \)^{- 4 h_L}  \, .  \label{eq:sumaribitrary} 
\end{align}
%

%%%%%%%%%%%%%%%%%%%%%%%%%%%%%%%%%%%%%%%%%%%%%%%%%%%%%%%%%%%%%%%%%%%%%%%%%
%%%%%%%%%%%%%%%%%%%%%%%%%%%%%%%%%%%%%%%%%%%%%%%%%%%%%%%%%%%%%%%%%%%%%%%%%

%%%%%%%%%%%%%%%%%%%%%%%%%%%%%%%%%%%%%%%%%%%%%%%%%%%%%%%%%%%%%%%%%%%%%%%%%%
%%%%%%%%%%%%%%%%%%%%%%%%%%%%%%%%%%%%%%%%%%%%%%%%%%%%%%%%%%%%%%%%%%%%%%%%%%


\begin{thebibliography}{99}


 %\cite{Fitzpatrick:2014vua}
\bibitem{Fitzpatrick:2014vua}
  A.~L.~Fitzpatrick, J.~Kaplan and M.~T.~Walters,
``Universality of Long-Distance AdS Physics from the CFT Bootstrap,''
  JHEP {\bf 1408} (2014) 145
    doi:10.1007/JHEP08(2014)145
 [\arXiv{arXiv:1403.6829} [hep-th]].

%\cite{Fitzpatrick:2015zha}
\bibitem{Fitzpatrick:2015zha}
  A.~L.~Fitzpatrick, J.~Kaplan and M.~T.~Walters,
  ``Virasoro Conformal Blocks and Thermality from Classical Background Fields,''
  JHEP {\bf 1511} (2015) 200
  doi:10.1007/JHEP11(2015)200
  [\arXiv{arXiv:1501.05315} [hep-th]].

%\cite{Hartman:2013mia}
\bibitem{Hartman:2013mia}
  T.~Hartman,
  ``Entanglement Entropy at Large Central Charge,''
  \arXiv{arXiv:1303.6955} [hep-th].

%\cite{Asplund:2014coa}
\bibitem{Asplund:2014coa}
  C.~T.~Asplund, A.~Bernamonti, F.~Galli and T.~Hartman,
  ``Holographic Entanglement Entropy from 2d CFT: Heavy States and Local Quenches,''
  JHEP {\bf 1502} (2015) 171
  doi:10.1007/JHEP02(2015)171
  [\arXiv{arXiv:1410.1392} [hep-th]].
  
%\cite{Caputa:2014eta}
\bibitem{Caputa:2014eta}
  P.~Caputa, J.~Sim\'on, A.~Stikonas and T.~Takayanagi,
  ``Quantum Entanglement of Localized Excited States at Finite Temperature,''
  JHEP {\bf 1501} (2015) 102
  doi:10.1007/JHEP01(2015)102
  [\arXiv{arXiv:1410.2287} [hep-th]].
  
%\cite{Fitzpatrick:2016ive}
\bibitem{Fitzpatrick:2016ive}
  A.~L.~Fitzpatrick, J.~Kaplan, D.~Li and J.~Wang,
  ``On information loss in AdS$_{3}$/CFT$_{2}$,''
  JHEP {\bf 1605} (2016) 109
  doi:10.1007/JHEP05(2016)109
  [\arXiv{arXiv:1603.08925} [hep-th]].  
  
 %\cite{Fitzpatrick:2016mjq}
\bibitem{Fitzpatrick:2016mjq}
  A.~L.~Fitzpatrick and J.~Kaplan,
  ``On the Late-Time Behavior of Virasoro Blocks and a Classification of Semiclassical Saddles,''
  JHEP {\bf 1704} (2017) 072
  doi:10.1007/JHEP04(2017)072
  [\arXiv{arXiv:1609.07153} [hep-th]].

%\cite{Galliani:2016cai}
\bibitem{Galliani:2016cai}
  A.~Galliani, S.~Giusto, E.~Moscato and R.~Russo,
  ``Correlators at large c without information loss,''
  JHEP {\bf 1609} (2016) 065
  doi:10.1007/JHEP09(2016)065
  [\arXiv{arXiv:1606.01119} [hep-th]].
 
%\cite{Asplund:2015eha}
\bibitem{Asplund:2015eha}
  C.~T.~Asplund, A.~Bernamonti, F.~Galli and T.~Hartman,
  ``Entanglement Scrambling in 2d Conformal Field Theory,''
  JHEP {\bf 1509} (2015) 110
  doi:10.1007/JHEP09(2015)110
  [\arXiv{arXiv:1506.03772} [hep-th]].
  
%\cite{deBoer:2016bov}
\bibitem{deBoer:2016bov}
  J.~de Boer and D.~Engelhardt,
  ``Remarks on thermalization in 2D CFT,''
  Phys.\ Rev.\ D {\bf 94} (2016) no.12,  126019
  doi:10.1103/PhysRevD.94.126019
  [\arXiv{arXiv:1604.05327} [hep-th]].
  
%\cite{Roberts:2014ifa}
\bibitem{Roberts:2014ifa}
  D.~A.~Roberts and D.~Stanford,
  ``Two-dimensional conformal field theory and the butterfly effect,''
  Phys.\ Rev.\ Lett.\  {\bf 115} (2015) no.13,  131603
  doi:10.1103/PhysRevLett.115.131603
  [\arXiv{arXiv:1412.5123} [hep-th]].  
  
  %\cite{Maldacena:2015waa}
\bibitem{Maldacena:2015waa}
  J.~Maldacena, S.~H.~Shenker and D.~Stanford,
  ``A bound on chaos,''
  JHEP {\bf 1608} (2016) 106
  doi:10.1007/JHEP08(2016)106
  [\arXiv{arXiv:1503.01409} [hep-th]].
  
  %\cite{Perlmutter:2016pkf}
\bibitem{Perlmutter:2016pkf}
  E.~Perlmutter,
  ``Bounding the Space of Holographic CFTs with Chaos,''
  JHEP {\bf 1610} (2016) 069
  doi:10.1007/JHEP10(2016)069
  [\arXiv{arXiv:1602.08272} [hep-th]].
 
%\cite{Chang:2015qfa}
\bibitem{Chang:2015qfa}
  C.~M.~Chang and Y.~H.~Lin,
  ``Bootstrapping 2D CFTs in the Semiclassical Limit,''
  JHEP {\bf 1608} (2016) 056
  doi:10.1007/JHEP08(2016)056
  [\arXiv{arXiv:1510.02464} [hep-th]].
  
%\cite{Collier:2017shs}
\bibitem{Collier:2017shs}
  S.~Collier, P.~Kravchuk, Y.~H.~Lin and X.~Yin,
  ``Bootstrapping the Spectral Function: On the Uniqueness of Liouville and the Universality of BTZ,''
  \arXiv{arXiv:1702.00423} [hep-th].
  
 %\cite{Menotti:2004xz}
\bibitem{Menotti:2004xz}
  P.~Menotti and G.~Vajente,
  ``Semiclassical and quantum Liouville theory on the sphere,''
  Nucl.\ Phys.\ B {\bf 709} (2005) 465
  doi:10.1016/j.nuclphysb.2004.12.014
  [\arXiv{hep-th/0411003}].

%\cite{Menotti:2005fk}
\bibitem{Menotti:2005fk}
  P.~Menotti and E.~Tonni,
  ``Quantum Liouville theory on the pseudosphere with heavy charges,''
  Phys.\ Lett.\ B {\bf 633} (2006) 404
  doi:10.1016/j.physletb.2005.11.061
  [\arXiv{hep-th/0508240}].

%\cite{Menotti:2006gc}
\bibitem{Menotti:2006gc}
  P.~Menotti and E.~Tonni,
  ``Liouville field theory with heavy charges. I. The Pseudosphere,''
  JHEP {\bf 0606} (2006) 020
  doi:10.1088/1126-6708/2006/06/020
  [\arXiv{hep-th/0602206}].
    
%\cite{Menotti:2006tc}
\bibitem{Menotti:2006tc}
  P.~Menotti and E.~Tonni,
  ``Liouville field theory with heavy charges. II. The Conformal boundary case,''
  JHEP {\bf 0606} (2006) 022
  doi:10.1088/1126-6708/2006/06/022
  [\arXiv{hep-th/0602221}].
   
%\cite{Harlow:2011ny}
\bibitem{Harlow:2011ny}
  D.~Harlow, J.~Maltz and E.~Witten,
  ``Analytic Continuation of Liouville Theory,''
  JHEP {\bf 1112} (2011) 071
    doi:10.1007/JHEP12(2011)071
  [\arXiv{arXiv:1108.4417} [hep-th]].

 %\cite{Zamolodchikov:1995aa}
\bibitem{Zamolodchikov:1995aa}
  A.~B.~Zamolodchikov and A.~B.~Zamolodchikov,
 ``Structure constants and conformal bootstrap in Liouville field theory,''
  Nucl.\ Phys.\ B {\bf 477} (1996) 577
    doi:10.1016/0550-3213(96)00351-3
  [\arXiv{hep-th/9506136}].

%\cite{Hadasz:2006rb}
\bibitem{Hadasz:2006rb}
  L.~Hadasz and Z.~Jaskolski,
  ``Liouville theory and uniformization of four-punctured sphere,''
  J.\ Math.\ Phys.\  {\bf 47} (2006) 082304
  doi:10.1063/1.2234272
  [\arXiv{hep-th/0604187}].  
  
%\cite{Krasnov:2000ia}
\bibitem{Krasnov:2000ia}
  K.~Krasnov,
  ``3-D gravity, point particles and Liouville theory,''
  Class.\ Quant.\ Grav.\  {\bf 18} (2001) 1291
  doi:10.1088/0264-9381/18/7/311
  [\arXiv{hep-th/0008253}].

%\cite{Gary:2009ae}
\bibitem{Gary:2009ae}
  M.~Gary, S.~B.~Giddings and J.~Penedones,
  ``Local bulk S-matrix elements and CFT singularities,''
  Phys.\ Rev.\ D {\bf 80} (2009) 085005
  doi:10.1103/PhysRevD.80.085005
  [\arXiv{arXiv:0903.4437} [hep-th]].
  
%\cite{Heemskerk:2009pn}
\bibitem{Heemskerk:2009pn}
  I.~Heemskerk, J.~Penedones, J.~Polchinski and J.~Sully,
  ``Holography from Conformal Field Theory,''
  JHEP {\bf 0910} (2009) 079
  doi:10.1088/1126-6708/2009/10/079
[\arXiv{arXiv:0907.0151} [hep-th]].

%\cite{ZZ}
\bibitem{ZZ}
A.~B.~Zamolodchikov and A.~B.~Zamolodchikov,
\textit{http://qft.itp.ac.ru/ZZ.pdf}.
     
%\cite{Seiberg:1990eb}
\bibitem{Seiberg:1990eb}
  N.~Seiberg,
  ``Notes on quantum Liouville theory and quantum gravity,''
  Prog.\ Theor.\ Phys.\ Suppl.\  {\bf 102} (1990) 319.
    doi:10.1143/PTPS.102.319

%\cite{Teschner:2001rv}
\bibitem{Teschner:2001rv}
  J.~Teschner,
  ``Liouville theory revisited,''
  Class.\ Quant.\ Grav.\  {\bf 18} (2001) R153
  doi:10.1088/0264-9381/18/23/201
  [\arXiv{hep-th/0104158}].
  
%\cite{Dorn:1994xn}
\bibitem{Dorn:1994xn}
  H.~Dorn and H.~J.~Otto,
  ``Two and three point functions in Liouville theory,''
  Nucl.\ Phys.\ B {\bf 429} (1994) 375
    doi:10.1016/0550-3213(94)00352-1
  [\arXiv{hep-th/9403141}].

%\cite{Beccaria:2015shq}
\bibitem{Beccaria:2015shq}
  M.~Beccaria, A.~Fachechi and G.~Macorini,
  ``Virasoro vacuum block at next-to-leading order in the heavy-light limit,''
  JHEP {\bf 1602} (2016) 072
  doi:10.1007/JHEP02(2016)072
  [\arXiv{arXiv:1511.05452} [hep-th]].
 
%\cite{Fitzpatrick:2015dlt}
\bibitem{Fitzpatrick:2015dlt}
  A.~L.~Fitzpatrick and J.~Kaplan,
  ``Conformal Blocks Beyond the Semi-Classical Limit,''
  JHEP {\bf 1605} (2016) 075
  doi:10.1007/JHEP05(2016)075
  [\arXiv{arXiv:1512.03052} [hep-th]].
  
\bibitem{Appolonius}
Ecyclopaedia Brittanica, https://www.britannica.com/biography/Apollonius-of-Perga;  Wikipedia, https://en.wikipedia.org/wiki/Circles\_of\_Apollonius.
               
%\cite{Kulaxizi:2014nma}
\bibitem{Kulaxizi:2014nma}
  M.~Kulaxizi, A.~Parnachev and G.~Policastro,
  ``Conformal Blocks and Negativity at Large Central Charge,''
  JHEP {\bf 1409} (2014) 010
  doi:10.1007/JHEP09(2014)010
  [\arXiv{arXiv:1407.0324} [hep-th]].
  
 %\cite{Cardy:2007mb}
\bibitem{Cardy:2007mb}
  J.~L.~Cardy, O.~A.~Castro-Alvaredo and B.~Doyon,
  ``Form factors of branch-point twist fields in quantum integrable models and entanglement entropy,''
  J.\ Statist.\ Phys.\  {\bf 130} (2008) 129
  doi:10.1007/s10955-007-9422-x
  [\arXiv{arXiv:0706.3384} [hep-th]].
  
  %\cite{Dolan:2000ut}
\bibitem{Dolan:2000ut}
  F.~A.~Dolan and H.~Osborn,
  ``Conformal four point functions and the operator product expansion,''
  Nucl.\ Phys.\ B {\bf 599} (2001) 459
  doi:10.1016/S0550-3213(01)00013-X
[\arXiv{hep-th/0011040}].


\end{thebibliography}
\end{document}